# Injectable Bubbles for Physiological Pressure Measurement

---

**Prashant Pandey**

**Brasenose College**

**Supervisors:** Professor Eleanor Stride and Professor Robin Cleveland

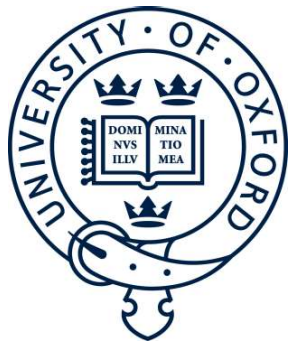

**Department of Engineering Science, University of Oxford**

2015 – 2016

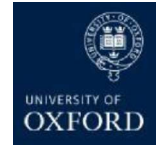

**FINAL HONOUR SCHOOL OF ENGINEERING SCIENCE**
**DECLARATION OF AUTHORSHIP**

You should complete this certificate. It should be bound into your fourth year project report, immediately after your title page. **Three copies** of the report should be submitted to the Chairman of Examiners for your Honour School, c/o Clerk of the Schools, examination Schools, High Street, Oxford.

**Name (in capitals):** PRASHANT PANDEY          **Supervisor:** ELEANOR STRIDE

**College (in capitals):** BRASENOSE

**Title of project (in capitals):** INJECTABLE BUBBLES FOR PHYSIOLOGICAL PRESSURE MEASUREMENT

**Page count (excluding risk and COSHH assessments:** 50

*Please tick to confirm the following:*

I have read and understood the University's disciplinary regulations concerning conduct in examinations and, in particular, the regulations on plagiarism (*Essential Information for Students. The Proctors' and Assessor's Memorandum*, Section 9.6; also available at www.admin.ox.ac.uk/proctors/info/pam/section9.shtml). ✓

I have read and understood the Education Committee's information and guidance on academic good practice and plagiarism at www.admin.ox.ac.uk/edc/goodpractice. ✓

The project report I am submitting is entirely my own work except where otherwise indicated. ✓

It has not been submitted, either partially or in full, for another Honour School or qualification of this University (except where the Special Regulations for the subject permit this), or for a qualification at any other institution. ✓

I have clearly indicated the presence of all material I have quoted from other sources, including any diagrams, charts, tables or graphs. ✓

I have clearly indicated the presence of all paraphrased material with appropriate references. ✓

I have acknowledged appropriately any assistance I have received in addition to that provided by my supervisor. ✓

I have not copied from the work of any other candidate. ✓

I have not used the services of any agency providing specimen, model or ghostwritten work in the preparation of this project. (See also section 2.4 of Statute XI on University Discipline under which members of the University are prohibited from providing material of this nature for candidates in examinations at this University or elsewhere: http://www.admin.ox.ac.uk/statutes/352-051a.shtml# Toc28142348.) ✓

The project report (excluding risk and COSHH assessments) does not exceed 50 pages (including all diagrams, photographs, references and appendices). ✓

I agree to retain an electronic copy of this work until the publication of my final examination result, except where submission in hand-written format is permitted. ✓

I agree to make any such electronic copy available to the examiners should it be necessary to confirm my word count or to check for plagiarism. ✓

**Candidate's signature:**          **Date:**    15–04–2016

# Acknowledgements

This project would not have been possible without the contribution of various people, whom I wish to thank. I am indebted to my research supervisors, Professor Eleanor Stride and Professor Robin Cleveland, who provided invaluable guidance and advice throughout the project.

I would also like to thank Dr Yesna Yildiz for teaching me how to use the Verasonics® equipment, and her enthusiastic support for my experimental work. I want to express my gratitude to Steve Bian for allowing me to use his experimental equipment, and patiently showing me how to make the most of the laboratory's facilities.

Finally, I am grateful to Jovi Wong for inspiring my passion for research, and for her loving support.

# Abstract

Microbubbles – used as contrast agents in ultrasound imaging – are important tools in biomedical research, having been used together with ultrasound to develop significant diagnostic and therapeutic techniques. It has been suggested that the dynamic behaviour of microbubbles is dependent on the surrounding fluid's ambient (hydrostatic) pressure, and the potential to non-invasively determine blood pressure has numerous medical applications. To study this dependence, a computational mathematical model was created based on Marmottant's dynamic model of a microbubble [1]. A pulse inversion (PI) protocol was incorporated into the model to emphasize the nonlinear behaviour of the microbubble's response. Using the PI signal energy as a metric, it can be shown that it is feasible to observe a change in a population of microbubbles when the ambient pressure is increased by as little as 7.5 mmHg (1000 Pa – a 1% change in atmospheric pressure). The mathematical model was also used to assess the sensitivity of the microbubbles to undesirable changes in parameters other than the ambient pressure. It found that a variation in the microbubble's initial radius and surface tension would cause the most significant changes in signal energy and hence pose a risk to ambient pressure measurements.

   To test the practicality of detecting a change in the dynamic behaviour of microbubbles, *in vitro* experiments were designed and carried out using clinically available contrast agent with two different ultrasound systems. The experiments, while possessing certain limitations, confirmed that there is a change in microbubbles' dynamic behaviour when the ambient pressure is varied, in cases by as little as 7.36 mmHg (981 Pa – a 0.98% change in atmospheric pressure). The experimental results establish a proof-of-principle that future experimental work can build upon to verify the mathematical model, and hence aid in developing a non-invasive blood pressure measurement procedure.

# Contents

# 1. Introduction

## 1.1 Background and Motivation

Ultrasound imaging for medical diagnosis has grown enormously for over sixty years. It has become ubiquitous due to its safety, low-cost, portability of the equipment, and the speed with which imaging is done. The operation of medical ultrasound relies on the propagation of sound from a transducer to a location within the patient's body, and the reception of ultrasonic echoes from that region. The physics of ultrasound are no different from those of sound: ultrasound waves are pressure waves with frequencies above the human hearing limit of 20 kHz. In diagnostic imaging ultrasound transducers emit a 1 cycle pulse between 20 kHz – 16 MHz depending on the application [2]. As the pulse propagates through the body it is reflected (scattered) by acoustic interfaces. The arrival time of the reflection is used to determine the depth of the interface in the body, and the amplitude determines the strength of the scattering. This information is used to construct an image of the target. Interfaces with a large acoustic impedance difference produce strong echoes and hence good imaging contrast, whereas boundaries with small impedance differences do not.

A significant body of research has focused on the development of ultrasound contrast agents. These are gas filled microbubbles which typically have a diameter between 1 and 8 µm and are coated with a stabilising shell [3]. As the gas within microbubbles is encapsulated and highly compressible, the microbubbles undergo significant volumetric oscillations under the influence of an ultrasound field. These time-varying oscillations make microbubbles strong scatterers of ultrasound. In addition, microbubbles oscillate nonlinearly whereas the surrounding tissue environment mainly responds to the ultrasound signal in a linear manner. Hence, microbubbles are often injected into the circulatory system to enhance the contrast of ultrasound diagnostic images. Microbubbles have also been studied for their therapeutic potential, such as for targeted drug delivery and tissue ablation [2].

Analysis of the dynamic behaviour of microbubbles has suggested that their behaviour is sensitive to the ambient pressure of the surrounding fluid, an idea first proposed by Fairbank in 1977 [4]. Hence, it has been proposed that microbubbles can be used to non-invasively probe the fluid pressure within a patient. **The aim of this project is to examine the feasibility of this proposal.**

| Site | | Normal Pressure Range | |
|---|---|---|---|
| | | in mmHg | in kPa |
| Systemic arterial pressure | Systolic | 90 – 120 | 12 – 16 |
| | Diastolic | 60 – 79 | 8 – 11 |
| Central venous pressure | | 3 – 8 | 0.4 – 1.1 |
| Right ventricular pressure | Systolic | 15 – 30 | 2 – 4 |
| | Diastolic | 3 – 8 | 0.4 – 1.1 |
| Left ventricular pressure | Systolic | 100 – 140 | 13 – 19 |
| | Diastolic | 3 – 12 | 0.4 – 1.6 |
| Pulmonary artery pressure | Systolic | 15 – 30 | 2 – 4 |
| | Diastolic | 4 – 12 | 0.5 – 1.6 |
| Pulmonary vein pressure | | 2 – 15 | 0.3 – 2 |

**Table 1.1:** *Values of typical blood pressures at some of the sites in the circulatory system. The mean pressure values and normal ranges of each site vary significantly* [27]. *750 mmHg is equivalent to 100 kPa.*

The primary application of physiological fluid pressure measurement is blood pressure measurement. Blood pressure is a key indicator used by physicians to monitor the health of a patient, and to detect the onset of many diseases. With the exception of arterial blood pressure in systemic circulation, determining blood pressure in the body is not a trivial task. For instance to continuously monitor beat-by-beat blood pressure in non-systemic circulation, a catheter must be placed in the appropriate vessel or organ. This is a complex invasive procedure and increases the risk of harm to the patient, such as through infection [5]. Additionally, automated oscillatory blood pressure measurement systems used clinically are calibrated to measure only the systolic and diastolic arterial blood pressures. This introduces a limitation as the pressure of other components of the circulatory system cannot be non-invasively monitored, despite their importance. Table 1.1 summarises a range of blood pressure measurement sites with particular clinical relevance. It is evident that both the normal pressure values and typical ranges of the normal values vary significantly. As such, a method of non-invasively and continuously monitoring blood pressure anywhere in the circulatory system would greatly improve diagnostic procedures in medicine.

Furthermore, there are many areas of the body which would benefit from non-invasive pressure measurement of fluids other than blood. One example is the cerebrospinal fluid (CSF), which surrounds the brain. Hydrocephalus is a condition in which an excessive build-up of CSF exerts an abnormally high pressure on the brain and skull. Current diagnostic methods involve using Magnetic Resonance Imaging (MRI), which is a comparatively long and costly procedure. Moreover, physicians have to place a catheter directly into a ventricle in order to directly measure intracranial pressure, which is invasive and can raise complications for the patient [6].

There is thus considerable demand for a method for non-invasive fluid pressure measurements in medicine.

## 1.2 Review of Past Work

The concept of using ultrasound imaging to determine fluid pressure has been previously used in 1984 to measure the blood pressure in the right ventricle of the heart [7]. Research has also been conducted to determine how microbubble contrast agents can be used with ultrasound to probe ambient fluid pressure. Fairbank (1977) proposed that a change in pressure can be detected by the corresponding shift in resonant frequency of a population of microbubbles [4]. However this study was limited as the sensitivity of the frequency shift to a change in pressure was very low: roughly a 10 kHz change in resonance frequency per 150 mmHg (20 kPa) of change in fluid pressure. Furthermore it was difficult to characterise the response of the microbubbles due to their large size distribution. A proportion of research that followed has also pursued the idea of observing a resonant frequency shift caused by increasing ambient pressure – Ishihara (1988) [8], Tremblay-Darveau (2011) [5].

An alternative approach is to detect changes in the frequency content – particularly the subharmonic frequency – of the back-scattered pressure signal to determine changes in the ambient pressure, as proposed by Shi (1998) [9]. The subharmonic response has been further studied by Frinking [10], Katiyar and Sarkar [11], Andersen [12], and Mobadersany [13]. Particularly, Dave *et al.* (2012) [14] validated a subharmonic aided pressure estimation system both *in vivo* and *in vitro* to find that the non-invasive measurements produced errors as low as 2.84 mmHg (380 Pa), when compared to reference measurements taken by an invasive pressure catheter. However, such a technique is limited as it depends entirely on the subharmonic frequency band and thus requires extensive calibration of the ultrasound transducer to maximise sensitivity to the subharmonic response, as reported by the authors.

A common limitation of most of the studies cited above has been the lack of sensitivity for clinical applications. From the values in Table 1.1, it can be seen that at some sites a sensitivity as low as 5 mmHg would be required in order to determine abnormal blood pressures. Other comprehensive studies in cardiovascular medicine show that the ability to measure a change of at least 10 − 15 mmHg will have a significant clinical impact − particularly in determining venous pressures and resulting pathologies [15].

## 1.3  Project Objectives

The primary aim of this project is to develop a mathematical and computational model which will be able to predict the response of a population of microbubbles due to changes in the ambient pressure of the fluid surrounding the microbubbles. Specifically this project aims to:

- Numerically model the dynamic behaviour of a microbubble in response to megahertz ultrasound (Chapter 2, sections 2.1 and 2.2).

- Model the response of a microbubble to a change in fluid ambient pressure (Chapter 2, sections 2.3 and 2.4).

- Perform a sensitivity analysis of the numerical model to determine the effects of changes in parameters other than the ambient pressure (Chapter 2, section 2.5).

- Model the behaviour of a population of microbubbles whose properties are not homogenous (Chapter 2, section 2.6).

- Design and carry out experimental work to demonstrate proof-of-principle of the feasibility of detecting a change in signal energy of clinically available contrast agent when ambient pressure is varied (Chapter 3).

- Present limitations of the numerical and experimental models, together with suggestions for their improvements to aid future research (Chapter 4).

To be considered clinically significant it is necessary that a change in ambient pressure of at least 10 mmHg (1300 Pa), if not less, can be detected.

# 2. Theory and Mathematical Model

## 2.1 Dynamic Response of a Free Microbubble

All microbubble contrast agents in clinical use have a surface coating which stabilises the particles during oscillation, and slows down the rate of diffusion of gas out of the bubble. However, when the behaviour of bubbles was first studied by Lord Rayleigh in the early twentieth century, the particles were considered to be 'free' bubbles without any coating over the gas/liquid interface. In this section a model of a free bubble will be considered, based on work by several authors over the twentieth century [16]. In the next section, the model will be extended to consider the effect of a surface coating.

If the radius of the microbubble is much smaller than the wavelength of the applied ultrasound field, only spherically symmetric volume oscillations are expected. Thus the bubble can be modelled as a gas sphere in an infinite liquid under the influence of an applied pressure field, whose radius varies with time. At a radius $r$ in the fluid, conservation of mass for an incompressible flow requires $4\pi r^2 \dot{r} = 4\pi R^2 \dot{R}$ and hence $\dot{r}/\dot{R} = R^2/r^2$. The kinetic energy of the surrounding fluid due to the oscillating bubble can then be expressed as:

$$E_K = \int_R^\infty \frac{1}{2} \rho_L \vec{v}^2 dV = 2\rho_L \pi \int_R^\infty r^2 \dot{r}^2 dr = 2\rho_L \pi R^4 \dot{R}^2 \int_R^\infty r^{-2} dr \qquad 2.1$$

Where $dV = 4\pi r^2 dr$ is a volume element, $\rho_L$ is the fluid density, where $\dot{r}$ is the fluid velocity in the radial direction. Ignoring the losses due to fluid viscosity, the kinetic energy of the fluid must be equal to the change in work done by the pressure from the initial position of the bubble wall $R_0$ to the current position $R$:

$$E_K = \Delta W$$

$$2\rho_L \pi R^4 \dot{R}^2 \int_R^\infty r^{-2} dr = \int_{R_0}^R (P_L(R) - P_\infty) 4\pi R^2 dR$$

$$2\pi \rho_L \dot{R}^2 R^3 = \int_{R_0}^R (P_L(R) - P_\infty) 4\pi R^2 dR$$

Differentiating with respect to $R$ yields:

$$2\pi\rho_L\left(2\ddot{R}R^3 + 3\dot{R}^2R^2\right) = (P_L - P_\infty)4\pi R^2$$

$$\frac{P_L - P_\infty}{\rho_L} = \frac{3}{2}\dot{R}^2 + \ddot{R}R \qquad\qquad 2.2$$

The effective pressure is the difference between the pressure in liquid at the bubble wall $P_L(R)$ and the pressure infinitely far from the bubble $P_\infty$. The pressure in the liquid $P_L$ plus the pressure due to surface tension (Laplace pressure) will equal the pressure within the bubble. The pressure within the bubble will be composed of the pressure of the gas $P_G$ and vapour $P_V$:

$$P_G + P_V = P_L + \frac{2\sigma}{R} \qquad\qquad 2.3$$

Where $\sigma$ is the surface tension of the air-water interface at the bubble membrane. At equilibrium, when there is no applied ultrasound field, the pressure of the fluid is $P_L = P_0$, and the bubble has an equilibrium radius $R = R_0$. Assuming $P_V$ is negligible, equation 2.3 then yields:

$$P_{G,e} = P_0 + \frac{2\sigma}{R_0} \qquad\qquad 2.4$$

The polytropic gas law states that $PV^k = constant$, where $k$ is the gas' polytropic exponent, and $V$ the volume of the gas. Therefore as the bubble oscillates the pressure of the gas varies as:

$$P_G = P_{G,e}\left(\frac{R_0}{R}\right)^{3k} \qquad\qquad 2.5$$

If the pressure remote to the bubble $P_\infty = P_0 + P_{ac}(t)$, where $P_0$ is the ambient (hydrostatic) pressure of the fluid and $P_{ac}(t)$ is the applied acoustic pressure field, then equation 2.2 becomes:

$$\rho_L\left(\frac{3}{2}\dot{R}^2 + \ddot{R}R\right) = \left(P_0 + \frac{2\sigma}{R_0}\right)\left(\frac{R_0}{R}\right)^{3k} - \frac{2\sigma}{R} - P_0 - P_{ac}(t) \qquad\qquad 2.6$$

The effect of viscosity was modelled by Plesset (1977) [17], and introduces losses in energy due to the viscosity $\mu_L$ of the fluid surrounding a bubble. Equation 2.6 then assumes the form of the Rayleigh-Plesset-Noltingk-Neppiras-Poritsky (RPNNP) equation:

$$\rho_L\left(\frac{3}{2}\dot{R}^2 + \ddot{R}R\right) = \left(P_0 + \frac{2\sigma}{R_0}\right)\left(\frac{R_0}{R}\right)^{3k} - \frac{2\sigma}{R} - \frac{4\mu_L\dot{R}}{R} - P_0 - P_{ac}(t) \qquad\qquad 2.7$$

$R, \dot{R}, \ddot{R}$ are the instantaneous displacement, velocity and acceleration of the bubble membrane respectively, and are functions of time.

## 2.2 Dynamic Response of a Coated Microbubble

Without surface coatings, microbubble contrast agents would not survive for sufficient time in the circulation to be useful. The gas inside an uncoated microbubble diffuses out to the surrounding fluid over milliseconds, resulting in the loss of acoustic contrast [5]. The predominant driving force is the large gas concentration gradient across the microbubble interface. After the introduction of surface coatings, it was observed that the behaviour of a microbubble in response to ultrasound does not follow the model as described in equation 2.7. Coated microbubbles have an attenuated response (lower amplitude radial oscillations) and a shifted resonant frequency in comparison to free microbubbles [5]. It is generally agreed that this change in dynamic behaviour is a result of the stiffness and viscosity of the shell layer around the bubble membrane.

Several models, based on the RPNNP equation, have been developed to describe the motion of coated contrast agents in response to ultrasound. These models differ depending on the material being considered for the shell, which can be made of phospholipids, polymers or proteins. This thesis will employ the model developed by Marmottant et al. in 2005 [1], which considers the coating of a microbubble to be formed of a phospholipid monolayer around the bubble and reflects the construction of most modern microbubble contrast agents. In particular, the Marmottant model has been cited as being very effective in accurately describing shelled microbubble behaviour in comparison to other models [5].

One key feature of the Marmottant model is that it captures the surface area dependency of the surface tension of a phospholipid coated microbubble, and hence the model describes the surface tension as a function of bubble radius. The changing nature of the surface tension is due to the concentration of lipid molecules increasing on the bubble surface as the bubble contracts (hence reducing surface area) and the opposite for when the bubble expands. Furthermore, the Marmottant model defines the microbubble as being in one of three different states: buckled, elastic or ruptured. In the elastic state, the surface tension value increases as the radius (and hence surface area) increases following a quadratic relationship. Below a certain threshold radius, the buckling radius ($R = R_{buck}$), the microbubble is in the buckled regime. In this regime the lipid layer buckles out of plane as there is not enough surface area to support the lipid molecules. This 'wrinkled' shell surface causes the surface tension to become negligible. Similarly above a threshold radius, the rupturing radius ($R = R_{rupt}$), the concentration of phospholipids will decrease and the surface tension will equal the value of the surrounding fluid. This complex behaviour is represented in the Marmottant model by a piecewise function:

$$\sigma(R) = \begin{cases} 0 & R \leq R_{buck} \\ \chi\left(\dfrac{R^2}{R_{buck}^2} - 1\right) & R_{buck} < R < R_{rupt} \\ \sigma_L & R \geq R_{rupt} \end{cases} \qquad 2.8$$

Where the constant $\chi$ is the elasticity of the shell, and $\sigma_L$ is the surface tension of the surrounding fluid (0.07 Nm$^{-1}$ for water). Equation 2.8 describes the changing nature of the surface tension, which was previously considered to be constant in equation 2.7. Both $R_{buck}$ and $R_{rupt}$ are also assumed to be constant, and are defined as:

$$R_{buck} = R_0 \left(\frac{\sigma(R_0)}{\chi} + 1\right)^{-\frac{1}{2}} \qquad 2.9$$

$$R_{rupt} = R_{buck} \left(1 + \frac{\sigma_L}{\chi}\right)^{\frac{1}{2}} \qquad 2.10$$

Where $\sigma(R_0)$ is the initial surface tension of the coated microbubble. The Marmottant model also incorporates a boundary condition containing a term for the viscosity of the shell as follows:

$$P_G = P_L + \frac{2\sigma(R)}{R} + \frac{4\mu_L \dot{R}}{R} + \frac{4\kappa_S \dot{R}}{R^2} \qquad 2.11$$

The right-most term embodies the effect of the shell viscosity, where $\kappa_S$ is the 'effective' viscosity of the shell. Together with the definition of surface tension in eqn. 2.8, this can be used to modify the RPNNP equation to give the Marmottant model:

$$\rho_L \left(\frac{3}{2}\dot{R}^2 + \ddot{R}R\right) = \left[P_0 + \frac{2\sigma(R_0)}{R_0}\right]\left(\frac{R_0}{R}\right)^{3k} - \frac{2\sigma(R)}{R} - \frac{4\mu_L \dot{R}}{R} - \frac{4\kappa_S \dot{R}}{R^2} - P_0 - P_{ac}(t) \qquad 2.12$$

Typical values of $\chi$ and $\kappa_S$ are of the order of 1 Nm$^{-1}$ and 10$^{-8}$ Nsm$^{-1}$ respectively, although these are difficult to measure in practice, and determination of these values rely on curve-fitting techniques [18]. Using the implementation of a Runge-Kutta solver in MATLAB (Release 2014b, The MathWorks, Inc., Natick, Massachusetts, United States), it is possible to numerically solve equations 2.7 and 2.12 to obtain the response of a microbubble to sinusoidal acoustic excitation (see Appendix A for the code). A comparison of the dynamic response of a free microbubble and a coated microbubble are presented in Figure 2.1. There are a number of differences between the two microbubbles. First, the coated response is much lower in amplitude than the free microbubble. Secondly, the amplitude of the coated microbubble is uniform, with a minimal transient response compared to the free microbubble. Furthermore, the frequency is slightly higher for the coated bubble – an indication of nonlinear effects and a shift in resonant frequency.

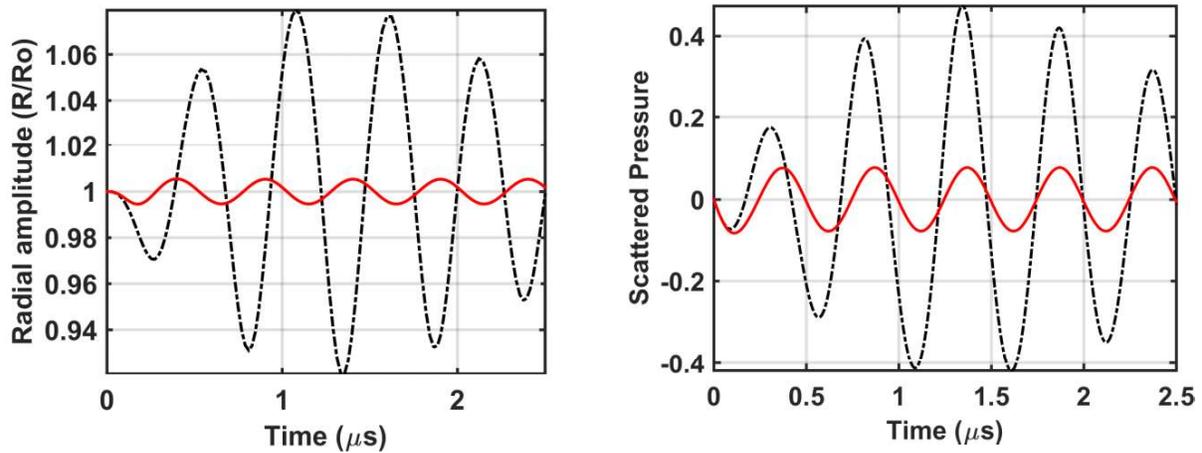

**Figure 2.1**: *The response of two microbubbles ($R_0 = 2\,\mu m$) to a sinusoidal ultrasound 5 cycle pulse with peak negative pressure 10 kPa and frequency of 2 MHz, at atmospheric ambient pressure. The black dashed-dotted line is the response of the free bubble, and the solid red line is a coated bubble described by the Marmottant model. **Left:** The instantaneous microbubble wall position (radial displacement) from the centre of the bubble for the duration of the applied signal. **Right:** Back-scattered pressure from the two microbubbles, at a distance $10 R_0$ from the bubble centre. Normalised with respect to the peak negative pressure of the ultrasound pulse.*

It is not trivial to directly measure the radial expansion and contraction of an actual microbubble under acoustic excitation. On the other hand, the back-scattered pressure can be readily measured using an ultrasonic transducer and is typically the most relevant quantity for imaging. This back-scattered pressure at a distance $d$ from the centre of the bubble is modelled theoretically as [19]:

$$P_{scat} = \rho_L \left[ \frac{1}{d} \left( R^2 \ddot{R} + 2R\dot{R}^2 \right) - \frac{R^4 \dot{R}^2}{2d^4} \right]$$

2.13

Hence, the scattered pressure signal can be simulated once the numerical solution to the dynamic model (eqn. 2.12) has been calculated. The scattered pressure of the two microbubble responses – non-dimensionalised with respect to the peak amplitude of the exciting ultrasound wave $|P_{ac}(t)|_{max}$ – is also presented in Figure 2.1. Similar differences are seen between the pressure waveforms as with the radial displacement waveforms.

## 2.3  Effect of Increasing Ambient Pressure (Overpressure)

A change in the ambient pressure of the fluid will have an effect on both the static (equilibrium) and dynamic behaviour of the microbubbles. Given that microbubbles are fabricated under atmospheric pressure ($10^5$ Pa), and the minimum possible physiological pressure will also equal atmospheric pressure, any possible change in physiological fluid pressure will have a compressive effect on the contrast agent. Hence it is important to first predict the change in static properties, $R_0$ and $\sigma(R_0)$, which will then determine the initial conditions of the dynamic model. To model the effect on the static behaviour (*i.e.* when no ultrasonic field is exciting the contrast agent), equation 2.4 can be used:

$$P_{G,e} = P_0 + \frac{2\sigma(R_0)}{R_0}$$

2.14

Considering the state before changing ambient pressure, denoted by subscript 1, and the state after changing pressure, denoted by subscript 2, and the ideal gas law (equation 2.5):

$$\left[ P_{01} + \frac{2\sigma(R_{01})}{R_{01}} \right] \left( \frac{R_{01}}{R_{02}} \right)^{3k} = \left[ P_{02} + \frac{2\sigma(R_{02})}{R_{02}} \right]$$

2.15

Where $P_{02} = P_{01} + P_{ov}$ and $P_{ov}$ is the additional fluid overpressure. Intuition from the ideal gas law will accurately predict that an increase in pressure around the microbubble will cause the compression of the gas within the bubble, hence decreasing the radius from $R_{01}$ to $R_{02}$. The solution to eqn. 2.15 (numerically solved in MATLAB – see Appendix A) is presented in Figure 2.2, and includes the effect of the changing surface tension as per the Marmottant model. The points of transition of the ruptured state to the elastic state and then to the buckled state are evident from the points at which the gradient changes substantially in both plots.

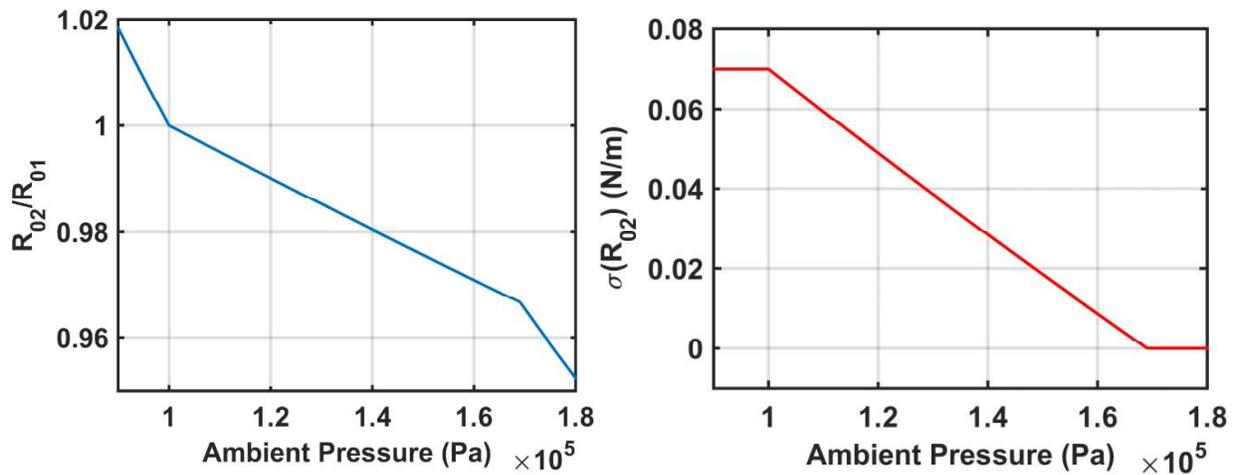

**Figure 2.2:** *The quasi-static effect of a change in ambient pressure on an air-filled microbubble ($k = 1.4$, $R_{01} = 3\ \mu m$) in water with initial pressure $P_{01} = 1\ kPa$.* **Left:** *The change in radius when pressure is varied above and below atmospheric pressure and* **Right:** *The resulting change in surface tension, initially at $\sigma(R_{01}) = 0.07\ Nm^{-1}$, decreasing to zero at approximately 170 kPa, which corresponds to buckling.*

## 2.4 Time-Frequency Signal Analysis

As reported in section 1.2, the frequency content of the dynamic response of a microbubble has been used as the main indicator of a change in the ambient environment. The frequency spectrum is obtained by applying a Fourier Transform to the time domain signal of the radial or pressure response of the microbubble. In a standard single pulse response, the frequency at which the largest peak appears is the fundamental frequency, and is approximately equal to the frequency $f$ of the applied acoustic signal $P_{ac}(t)$. Even if a sinusoidal pulse driven at a single frequency interacts with microbubbles, it is common for microbubbles to oscillate – and hence scatter pressure – at discrete frequencies other than the fundamental frequency. These are known as harmonics. Of particular importance for ultrasonic contrast agents are the second harmonic ($2f$) and the subharmonic ($0.5f$), which are at twice and half the fundamental frequency respectively. Other integer and half-integer multiples of the fundamental frequency are also present. The response contains these additional harmonics since the microbubble does not oscillate symmetrically: the volume of expansion does not equal the volume of compression. This is a manifestation of nonlinear behaviour, a property which further improves the contrast between microbubbles and their surrounding environment. This following section will examine the exploitation of nonlinear dynamic behaviour to determine a change in the ambient pressure.

The metric selected for the analysis of changes in the microbubble signal was the total signal energy, which takes into account the signal content across the entire frequency range. This is more robust than methods which examine the amplitude at particular harmonics. For short burst signals (about 5 µs) content at the subharmonic frequency has relatively low power compared to the fundamental content [20]. Furthermore, by definition, specific harmonic analysis does not take into account possible changes in frequency content outside the selected frequency, thus ignoring information which could be used to observe a change in the response of microbubbles. The energy of a signal is given by:

$$E_S = \int_{-\infty}^{\infty} |X(f)|^2 \, df \qquad\qquad 2.16$$

Where $X(f)$ is the complex frequency spectrum, given by the Fourier Transform of the time signal $x(t)$:

$$X(f) = \int_{-\infty}^{\infty} x(t)e^{-i2\pi ft} \, dt \qquad\qquad 2.17$$

Given the units of $x(t)$, it is possible to normalise $E_S$ by an impedance to determine the physical energy in Joules (or energy flux in Joules per unit area) – for instance the impedance in ohms for a voltage signal, or acoustic impedance ($Z_L = \rho_L c_L$, where $c_L$ is the speed of sound in the fluid) for a pressure signal. However as the impedance is considered to be constant, this approach is not necessary for calculating relative changes in signal energy.

In this project, a pulse inversion (PI) technique was used to improve the ability to detect changes in the nonlinear microbubble signal, by suppressing the scattered pressure from the surrounding fluid and tissue. PI is a common strategy in contrast enhanced ultrasound imaging [21], and requires two pulses to be transmitted consecutively, with the second one being inverted; see the left column of Figure 2.3. The resulting echoes from these two pulses are summed after they are received and any linear scattering sums to zero due to the inversion, resulting in a signal dependent on the nonlinear response of the microbubble – see Figure 2.3. Both pulse inversion and subsequent signal energy calculations were implemented in the MATLAB model of a single microbubble. This was achieved by simulating a short ultrasound pulses which excited the microbubble, and then modelling the dynamic response of the microbubble in the time and frequency domain. A short pulse was used in the experiments in order to maximise axial resolution and reduce the risk of destroying microbubbles. The pulse was used to excite a microbubble where $R_0$ and $\sigma(R_0)$ were dependent on a change in ambient pressure, as per equation 2.15. Figure 2.4 shows an example of the effect of changing the ambient pressure on the frequency response of a single microbubble. Under a pulse inversion protocol, the change in signal energy is significant. It is evident that linear scattering is negligible as there is little content present at the fundamental frequency, $f = 3$ MHz in the frequency spectrum.

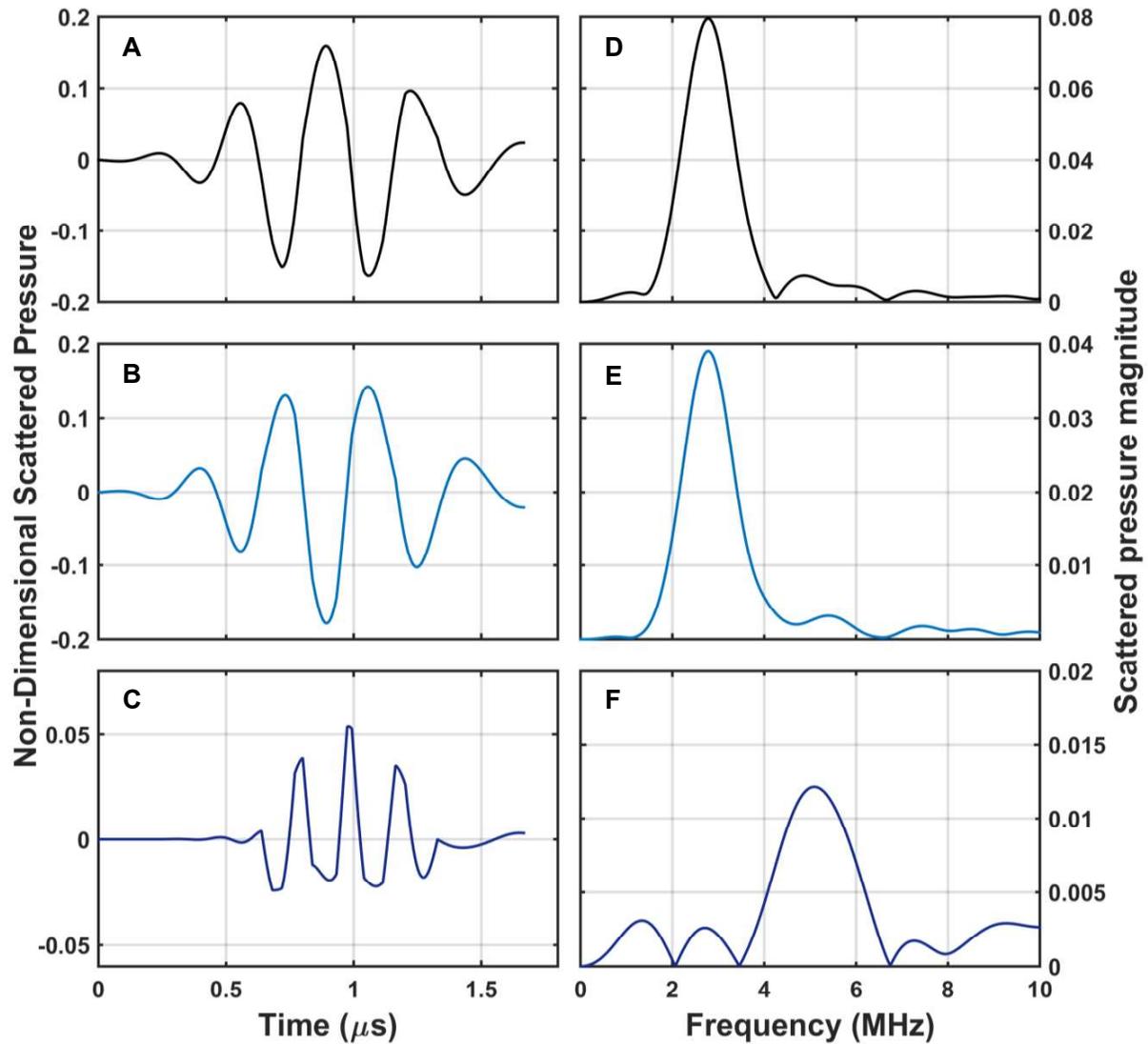

**Figure 2.3:** *The principles of pulse inversion (PI).* **Left column:** *Pressure time curves showing the back-scattered response of a single microbubble ($R_o = 3\ \mu m$) to 5 cycle pulses of frequency $f = 3$ MHz and peak pressure of 50 kPa. The response to a positive phase pulse is plotted in black (A), the response to a negative phase pulse ($180^o$ relative phase shift) is plotted in blue (B). The sum of these two responses forms the residual signal, plotted in navy (C).* **Right column:** *Frequency spectra of the respective time signals (using the Fast Fourier Transform). While the positive and negative phase responses (D, E) have the majority of content at the fundamental frequency (~3 MHz), for the residual signal the amplitude at the fundamental is relatively small compared to other frequencies (F), particularly the second harmonic: ~6 MHz. Thus, PI nullifies linear behaviour and emphasises nonlinearities.*

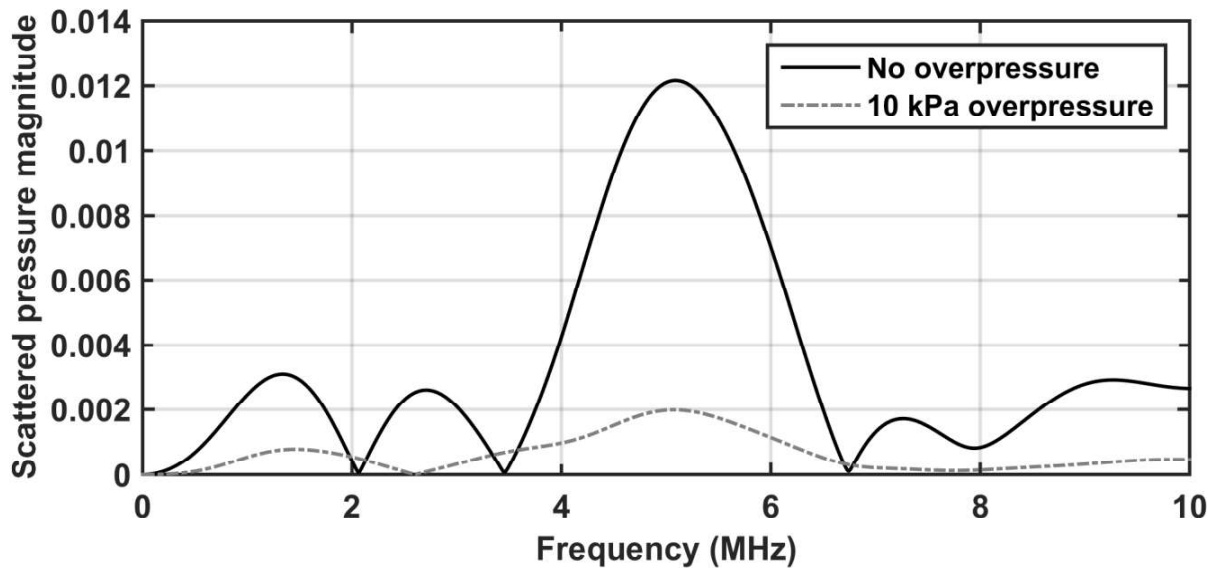

**Figure 2.4:** *Changes in hydrostatic pressure affect the frequency response of a microbubble, and hence the signal energy of the back scattered pressure. The grey dashed line is the frequency spectrum of the residual pressure response before any compression, and the black solid line is the spectrum after an application of 10 kPa (75 mmHg) overpressure. In this scenario, the compression yields an increase in signal energy by 3200%. $R_0 = 3\ \mu m$, $\sigma(R_0) = 0.035\ Nm^{-1}$, $\chi = 1\ Nm^{-1}$, $\kappa_S = 2.3 * 10^{-8}\ Nsm^{-1}$, under a PI protocol: 5 cycle pulse with frequency $f = 3\ MHz$, peak pressure $= 50\ kPa$.*

## 2.5  Sensitivity Analysis

It is evident from the RPNNP and Marmottant models that the dynamic response of microbubble contrast agents is dependent on several factors, including the parameter of primary interest in this study: the ambient pressure $P_0$. It is therefore necessary to carry out a sensitivity analysis using the model to determine whether it is feasible to detect a change in ambient pressure from a change in the scattered pressure signal, given expected variations in other parameters. This analysis should quantify the extent to which other factors have an effect on the response of the microbubble, and whether these can be considered negligible in comparison to the effect of changing hydrostatic pressure.

Table 2.1 summarises the relevant parameters from the model, as well as each parameter's chosen default value and the sensitivity of the signal energy to a $\pm$ 10% change of the parameter value. The default ambient pressure value was set to 100 kPa, as this will be the ambient pressure of fluid before pressurisation. The default values for fluid properties ($\mu_L$, $\rho_L$, $\sigma_L$) represent water, as the experiment will be carried out under water (see section 3). Microbubble properties ($\chi$, $\kappa_S$) were chosen from the most frequently cited values in literature, particularly from Marmottant [1] and Tremblay-Darveau [5]. The initial radius $R_0$ was set to that of a typical microbubble. The initial surface tension $\sigma(R_0)$ was chosen to be close to zero, such that an increase in ambient pressure would cause the microbubble to buckle and hence cause a noticeable change in behaviour and an increased sensitivity to increases in pressure. Values of $R_{buck}$ and $R_{rupt}$ were calculated from equations 2.9 and 2.10. The polytropic exponent was chosen to reflect air, as the analysis revealed that this value minimised sensitivity to changes in this parameter. This is also justified as although most clinically available contrast agents use a different gas core, such as perfluorocarbon ($k = 1.04$), the gas is substituted for air shortly after administration [22]. Furthermore, it is possible to fabricate in-house microbubbles with an air core that can be used for sufficient periods of time [19]. The peak pressure amplitude of the exciting pulse $|P_{ac}|_{max}$ was chosen such that microbubbles would respond nonlinearly, but also to minimise microbubble destruction. Finally, the frequency of the exciting pulses was chosen to be approximately twice the resonant frequency of the microbubble, as this is has been reported to increase nonlinear response and hence signal energy [10]. The sensitivity analysis was carried out in MATLAB, using the Marmottant model implementation, pulse inversion and signal energy calculations as described above. For each parameter, first the signal energy was calculated for a response using the parameter's default value – the default signal energy. Next, the default parameter value was increased by 10%, and the corresponding signal energy was calculated, followed by the same procedure for a decrease of the default value by 10%. Both of these 'perturbed' signal energies were compared to the default signal energy.

| Parameter (Description) | Default Value | Percentage change in signal energy after ±10% change in parameter | |
|---|---|---|---|
| | | +10% change | -10% change |
| $P_0$ (ambient/hydrostatic pressure) | 100 kPa | -98 | 160 |
| $R_0$ (initial bubble radius) | 3 μm | -36 | 89 |
| $\sigma(R_0)$ (initial bubble surface tension) | 0.0012 Nm$^{-1}$ | 3.6 | 0.5 |
| $\rho_L$ (fluid density) | 1000 kgm$^{-3}$ | -12 | 29 |
| $\mu_L$ (fluid viscosity) | 1 mPas | -3.3 | 1.4 |
| $\sigma_L$ (fluid surface tension) | 0.07 Nm$^{-1}$ | 0.0 | 0.0 |
| $\chi$ (shell elasticity) | 1 Nm$^{-1}$ | 19 | -16 |
| $\kappa_S$ (shell viscosity) | 2.3e-8 Nsm$^{-1}$ | 1.0 | 3.8 |
| $f$ (ultrasound frequency) | 3 MHz | -24 | 76 |
| $|P_{ac}|_{max}$ (peak pressure amplitude) | 50 kPa | 14 | -18 |
| $k$ (gas polytropic exponent) | 1.4 | n/a | -7.1 |

**Table 2.1:** *Details of the sensitivity analysis. The right-most column indicates the percentage change in signal energy $E_S$ as the parameter's value is perturbed. The first percentage indicates the energy change for a 10% increase in the parameter's value and the second percentage indicates the energy change for a 10% decrease in the parameter's value.*

The analysis reveals that the signal energy of the microbubble response was most sensitive to a change in ambient pressure, showing the largest change in signal energy compared to other parameters (a 98 percent change if the pressure increases by 10 kPa – or 75 mmHg, see Table 2.1). This is promising for the purpose of using microbubbles to detect a change in fluid pressure. However, other parameters have a possibility of introducing variability as they also cause a relatively large change in signal energy. Three parameters induced changes in signal energy of more than 20% - the initial radius $R_0$, ultrasound frequency $f$ and fluid density $\rho_L$. Therefore it is important to have accurate knowledge of those parameters such that the mathematical model can be used with certainty to detect a change in fluid pressure. Frequency $f$ is straightforward to maintain at a known value due to the high quality and tolerance of available signal generator equipment. Furthermore, the density $\rho_L$ of water is a known quantity, and can be assumed to remain constant due to the incompressible nature of the fluid, although this will not necessarily hold true in tissue. Changes in $\rho_L$ should be less than $\pm$ 3% to ensure that signal energy changes are less than $\pm$ 10%, determined through further analysis. The radius of the microbubble must also be maintained stringently, although this is not trivial given the typical size distributions of clinical contrast agents [23]. While it is expected for the change in ambient pressure to change the size of a microbubble, other factors affecting its size must be controlled. This includes the diffusion of gas out of the bubble and any changes in the ambient temperature. For the former, it is best to use a stiff and non-porous coating to minimise diffusion. Changes in ambient temperature can be controlled well *in vitro* using monitoring equipment, and *in vivo* given the constant temperature within the body. However, ultrasound waves dissipate energy as they propagate, which can cause the surrounding target environment to increase in temperature. Therefore, it is challenging to precisely control the change in microbubble radius under the influence of ultrasound. In addition, small changes in signal energy can occur due to changes in the other parameters and hence introduce uncertainty in energy measurements.

As a change in ambient pressure has a significant effect on the dynamic behaviour of a microbubble, the change in pulse inverted signal energy to a range of pressure values was calculated. This analysis is presented in Figure 2.5, which shows there is a large monotonic decrease of -56% in signal energy corresponding to a 3% increase in ambient pressure (equivalent to a 3 kPa or 22.5 mmHg increase). Moreover, a 1% increase in ambient pressure (corresponding to an increase of 1 kPa or 7.5 mmHg) resulted in a signal energy change of -11%. Therefore, there is significant potential in the ability to detect a pressure difference as small as 10 mmHg as discussed in section 1.3. Furthermore, under the aforementioned default parameter values, the ability to detect changes in pressure diminishes above an increase in pressure greater than 10% of $P_0$ (10 kPa or 75 mmHg). This value is greater than the upper limit of what is required in clinical applications.

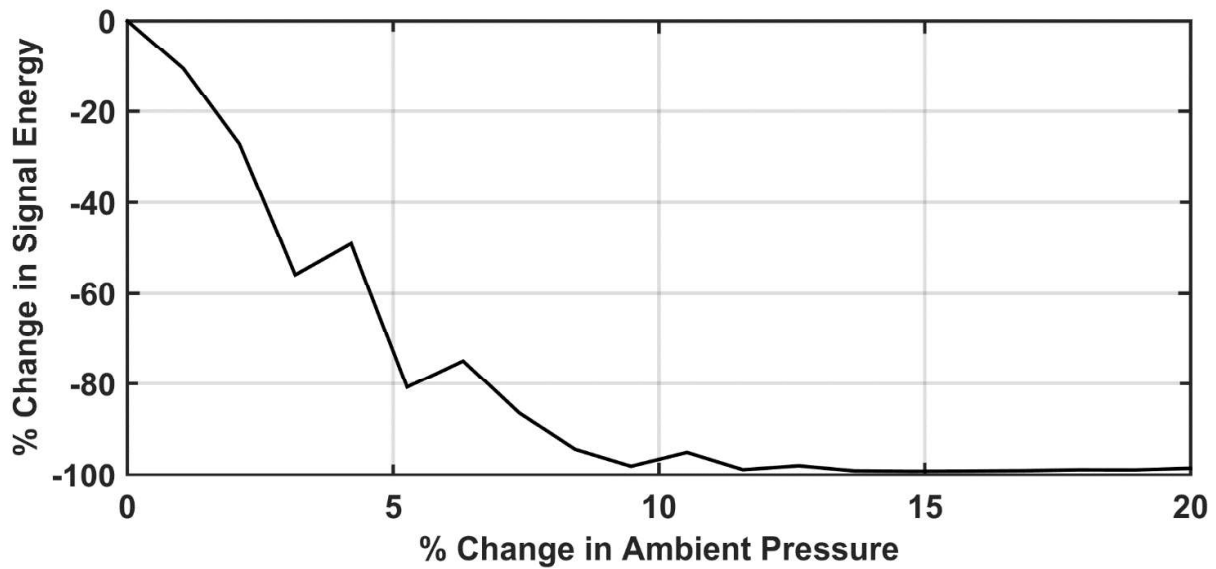

**Figure 2.5:** *The percentage change in signal energy of a single microbubble as the ambient pressure is increased from $P_0 = 100$ kPa. Model properties are taken from Table 2.1.*

## 2.6  Microbubble Population Analysis

The methods currently used to fabricate contrast agent particles result in a microbubble population with large variability in terms of bubble size and coating properties [23]. The previous analysis in this chapter has considered the response of a single microbubble only. This fails to account for the variance in dynamic response that will be inherent in a varying population as in a typical microbubble suspension. This section will present methods for extending the analysis to consider the effects of a population of bubbles, in order to more accurately represent a realistic scenario.

The central limit theorem provides a basis for approximating the distribution of microbubble properties for a relatively large number of microbubbles. Hence, properties such as the initial radii $R_0$ of microbubbles can be modelled to fit a normal distribution. Additionally, it has been suggested that the initial surface tension $\sigma(R_0)$ of the microbubbles are also distributed in a similar manner, and play an important role in the population's dynamics [23]. For a good approximation of the population, it is enough to sum the individual responses of each microbubble in the population under the assumption that each microbubble acts independently and linearly with respect to the others [24]. This is easily achievable computationally. To determine how many microbubbles should be included in the population, the focal volume and concentration of contrast agent particles should be considered. For the single element ultrasound transducer used in the experiments reported here, the focal volume is ellipsoidal (approximated to a cylinder) with radius 1 mm and length 3 mm, which gives a focal volume of 9.4 μl. This focal volume can be adjusted, by choosing a transducer with a different focal length or diameter [25]. The concentration of microbubbles is taken to be in the order of $10^4$ particles per ml, giving approximately 100 microbubbles in the focal volume of the transducer.

First the effect of variation in size was analysed. To model this, 100 microbubbles were considered with initial radii sampled from a normal distribution with mean radius $M_{R_0} = 3$ μm. The standard deviation of this distribution $\gamma_{R_0}$ was varied from $0$ μm $\leq \gamma_{R_o} \leq 1$ μm. The scattered pressure response to a pulse inversion scheme was calculated for each microbubble in the population, as in section 2.4. These responses were summed linearly to give the entire population's response, and thus allowing the total signal energy to be calculated. With each iteration the standard deviation of the size distribution was increased, in order to quantify the effect of a population's spread on the expected measurements of energy change.

The results – presented in Figure 2.6 – show that the trend in signal energy change is preserved despite variance in the bubble radii of a population. Additionally, a property of the varying populations is that any sudden changes in signal energy are not present – reflecting how the slightly varying responses of the individual microbubbles tend to smooth out the overall response. Conversely, in the monodisperse population each microbubble responds in exactly the same way, thus resulting in visibly sudden changes in signal energy. The population analysis reveals that even for a generously spread population with standard deviation $\gamma_{R_0} = 1$ μm, the response behaves in a similar manner to an ideal monodisperse population. The response monotonically decreases up to the calculated 20% increase in ambient pressure. However, this is not enough to critically evaluate the error resulting from the variation in radii of the population.

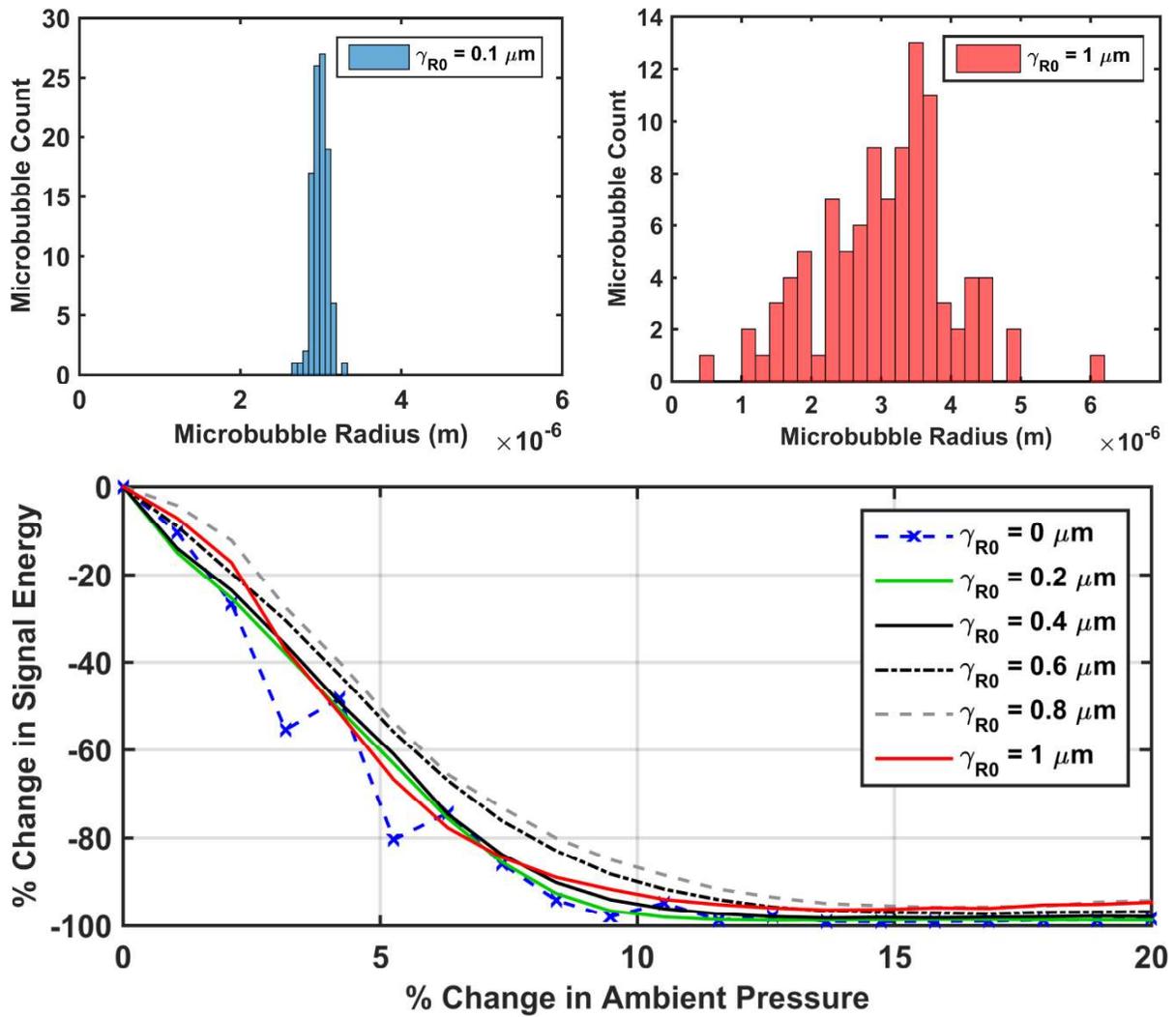

**Figure 2.6:** *Details of the population analysis considering microbubble radius as a normally distributed random variable, with mean $M_{R_0} = 3$ μm and 100 microbubbles.* **Top Left:** *Histogram of an example microbubble population with standard deviation $\gamma_{R_0} = 0.1$ μm.* **Top Right:** *Another randomly normally distributed population with $\gamma_{R_0} = 1$ μm.* **Bottom:** *The percentage change in signal energy in response to a change in ambient pressure for six different populations (including the ideal monodisperse population response with $\gamma_{R_0} = 0$ μm in blue dashes and crosses).*

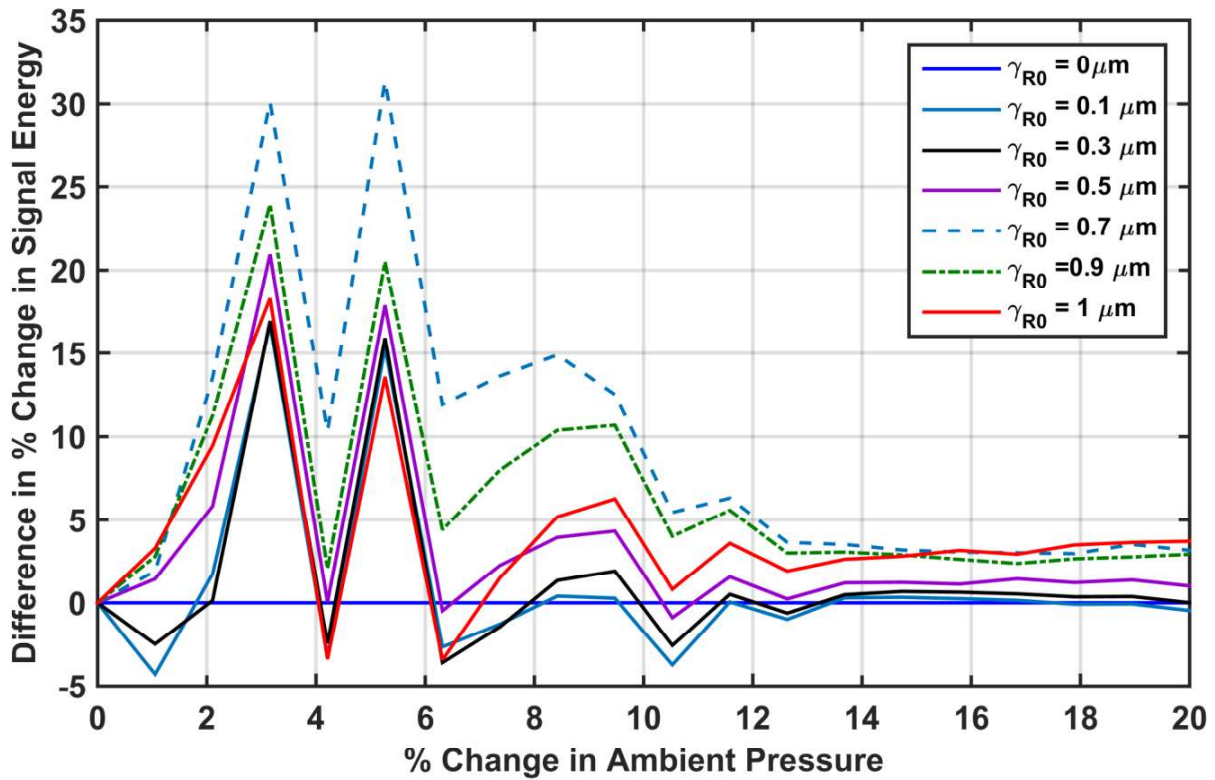

**Figure 2.7:** *Differences in percentage change in signal energy between the monodisperse population ($\gamma_{R_0} = 0\ \mu m$) and the other simulated populations. Signal energy peaks occurring at ambient pressure changes of 3.2% and 5.3% are ignored as they are a result of the ideal and unrealistic behaviour of a population where all microbubbles have the exact same property. Populations withstanding a spread up to $\gamma_{R_0} = 0.5\ \mu m$ show an error within the acceptable $\pm$ 5% boundary.*

Instead, the difference in the percentage change in signal energies between the varying populations and the monodisperse case was examined. This analysis is presented in Figure 2.7. Ignoring the large differences caused by the aforementioned sudden changes in energy of an ideal population, it can be shown that errors under $\pm$ 5% can be endured so long as the standard deviation of the population $\gamma_{R_0} \leq 0.5$ µm. This implies that 99% of the microbubbles in the population must have initial radii such that $1.5$ µm $\leq R_0 \leq 4.5$ µm, as per the properties of the Gaussian distribution. Once again, MATLAB was used to process this analysis –

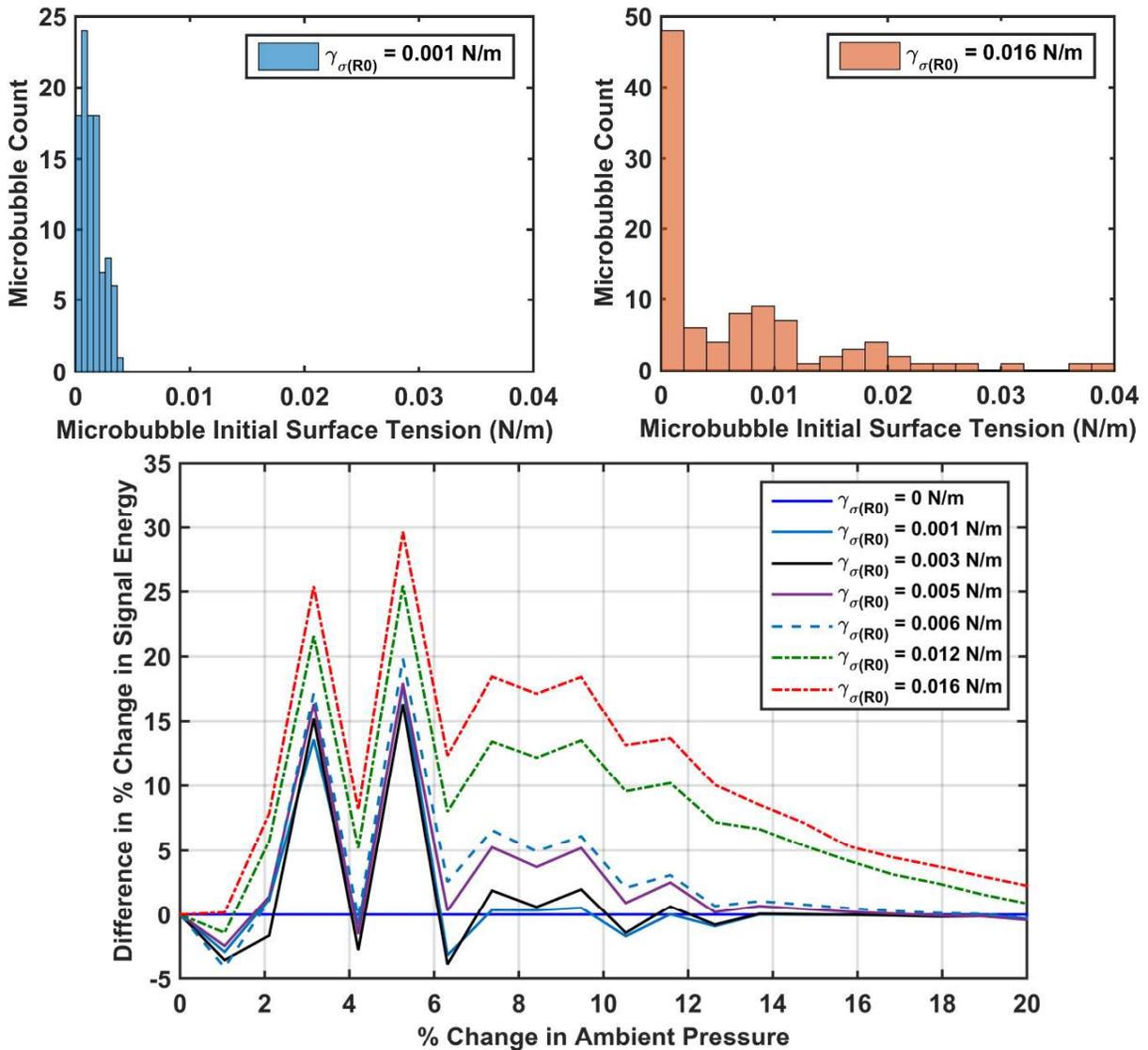

**Figure 2.8:** *Details of the population analysis considering variation in microbubble initial surface tension; population of 100 microbubbles with a corrected distribution such that $0 \leq \sigma(R_0) \leq 0.07\ Nm^{-1}$ for all microbubbles.* **Top Left:** *Histogram of an example of a correctly distributed microbubble population with Gaussian standard deviation $\gamma_{\sigma(R_0)} = 0.001\ Nm^{-1}$.* **Top Right:** *Histogram of another corrected population, more widely spread with Gaussian $\gamma_{\sigma(R_0)} = 0.016\ Nm^{-1}$.* **Bottom:** *Differences in percentage change in signal energy between the ideal population ($\gamma_{\sigma(R_0)} = 0\ Nm^{-1}$) and the other simulated populations. Populations withstanding a Gaussian spread up to $\gamma_{\sigma(R_0)} = 0.005\ \mu m$ show an error within the $\pm\ 5\%$ boundary.*

see Appendix A for the script.

Using a similar method, the effect of variance in initial surface tension $\sigma(R_0)$ in a microbubble population can be studied. However, a Gaussian distribution cannot be assumed in this scenario given the limits of surface tension prescribed by Marmottant in his model [1]. A modified Gaussian distribution was used with a mean $M_{\sigma(R_0)} = 0.0012\ Nm^{-1}$ (from Table 2.1), and varying standard deviations $0 \leq \gamma_{\sigma(R_0)} \leq 0.016\ Nm^{-1}$. Any surface tension values below 0 Nm$^{-1}$ (the minimum possible surface tension) were set to 0 Nm$^{-1}$ and any values above 0.07 Nm$^{-1}$ (the maximum possible value) were set to 0.07 Nm$^{-1}$. This reflects the nature of the buckled and ruptured states of a microbubble, as discussed in section 2.2. This process results in a population more accurately described by an exponential distribution – see Figure 2.8. Comparing signal energies to an ideal monodisperse population uses the same method as before; once again a varying population shows a much smoother response to an increase in ambient pressure than an ideal population where there is no variation in initial surface tension. The analysis reveals that an error of less than $\pm$ 5% in signal energy changes is possible if the Gaussian standard deviation used to initialise the population is such that $\gamma_{\sigma(R_0)} \leq 0.005$ Nm$^{-1}$ – as seen in Figure 2.8. The standard deviation of the actual population is different however, and can be determined by fitting a distribution to the population data. An exponential distribution fitted to the population reveals that the mean and standard deviation of such a population is equal to 0.0026 Nm$^{-1}$. While this does not exactly match the required population properties, such an analysis is useful in showing that a collection of microbubbles can withstand some variation in surface tension properties and still reliably indicate a change in hydrostatic pressure. In reality, compared to the microbubble size, it is difficult to measure and enforce the surface tension with precision during fabrication and before administration.

# 3. Experimental Models

## 3.1 Objectives

In Chapter 2, a numerical model was developed to allow the prediction of changes in signal energy of ultrasound echoes returned from a microbubble population when the ambient pressure was varied. The model predicted a detectable change in signal energy for pressure changes as small as 7.5 mmHg. Furthermore, to account for the unwanted effect of external factors on the signal energy, a detailed sensitivity analysis was performed. The sensitivity analysis showed that it is feasible for changes in ambient pressure to be detected despite possible changes in other parameters. The analysis was extended to consider the effects of a varying population of microbubbles, specifically size and surface tension variations. The limits of acceptable variation in these parameters were also quantified. For variation in size, the population had to be limited to a maximum standard deviation of 0.5 μm, and a standard deviation of 0.0026 $Nm^{-1}$ for a population variation in initial surface tension.

An ideal experimental model should enable these theoretical predictions to be tested. Within the scope of this project, however, the purpose of the experimental models was to determine the feasibility of detecting a change in signal energy caused by ambient pressure changes i.e. to demonstrate a proof-of-principle using currently available contrast agents and ultrasound systems.

## 3.2 Experiment Design and Set-up

The experimental design was guided by the physiological motivation of the project. To mimic the acoustic properties of tissue in the body, agar gel was used to prepare an imaging phantom consisting of a disc 40 mm in diameter and 9.6 mm in thickness. The phantom supported a hollow cylindrical channel (with a nominal 1.6 mm diameter) through its centre, representing an idealised blood vessel which would contain the suspension of microbubbles. A syringe pump was used to create a steady flow of the contrast agent solution through the channel at 0.2 μl/min. This would ensure that microbubbles within the focus of the transducer are replenished between pulses, as would happen with blood flow in the body. See Figure 3.1 for photographs of the phantom and channel.

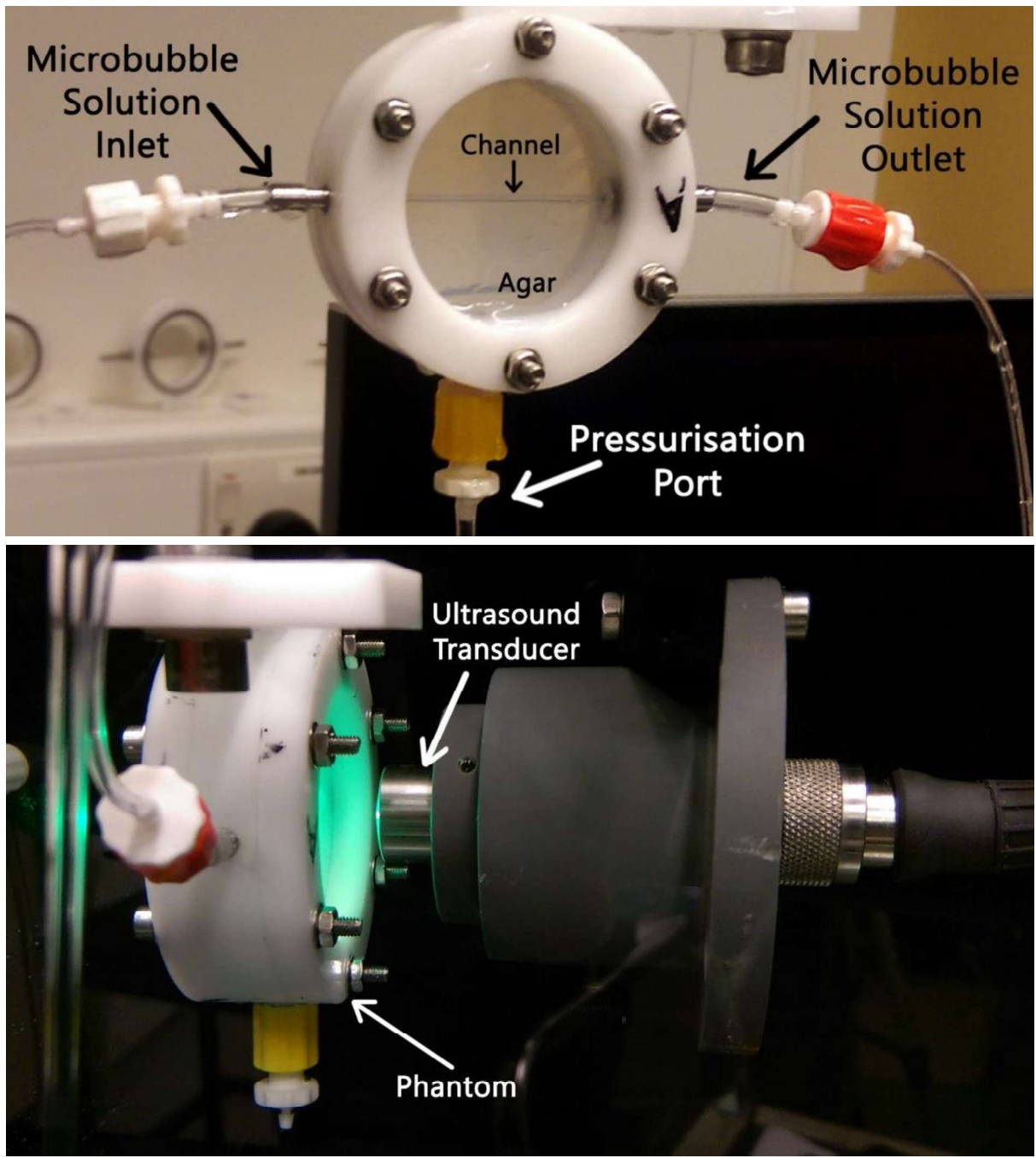

**Figure 3.1:**  *Photographs of the experimental setup.* **Top**: *The phantom was constructed using agar, and held between plastic retainers and two sheets of polyester film. The phantom supported a cylindrical channel in which the microbubbles flowed and were targeted by ultrasound.* **Bottom**: *The phantom and the transducer were secured and immersed in a tank of deionized water. The focus of the transducer was aligned with the centre of the channel. The transducer in this photograph is an Olympus immersion single element transducer with centre frequency of 2.25 MHz* [23].

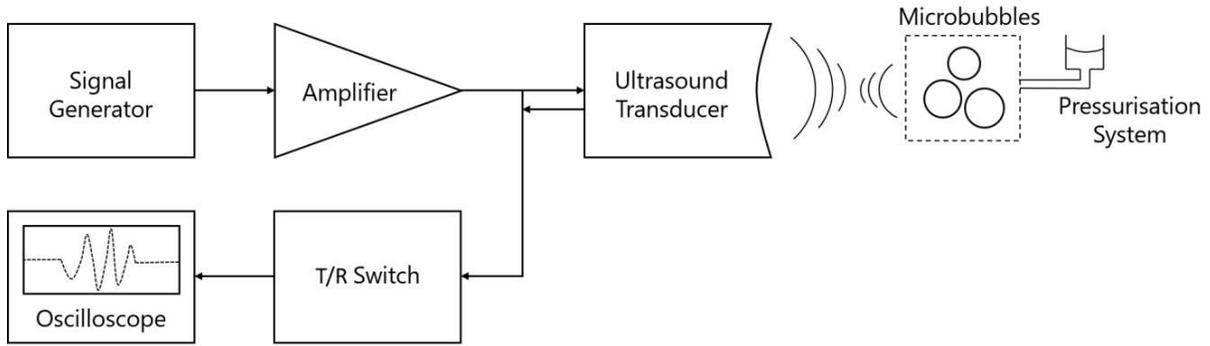

**Figure 3.2:** *Overview of the experimental setup. The input pulses were programmed and generated by the signal generator, which were then processed by a 55 dB voltage amplifier. This high voltage signal was transmitted by a single element ultrasound transducer to excite an ensemble of microbubbles within an agar phantom. Echoes from the microbubbles were received by the transducer. The transmit/receive (T/R) switch received both the 'transmitted' signal output from the amplifier and the subsequent echo signals 'received' by the transducer. These waveforms were viewed and stored by a digital oscilloscope. An independent pressurisation system increased and decreased the fluid pressure surrounding the microbubbles.*

A 50 ml syringe filled with water was used as the pressurisation mechanism. The syringe was connected to the phantom, such that the weight of the water caused the overpressure on the microbubbles within the channel. The exact hydrostatic pressure exerted by the water could be set by adjusting the height of the syringe relative to the phantom, given by:

$$P_0 = \rho_w g h \qquad\qquad 3.1$$

Where $P_0$ is the hydrostatic gauge pressure, $\rho_w$ is the density of water, and $h$ is the height difference between the water's surface in the syringe and the pressurisation port on the phantom holder. This method allowed control of the ambient pressure of the fluid surrounding the microbubbles.

Furthermore, the experimental model required the necessary equipment to ultrasonically excite the microbubbles, and to record the resulting response. This was achieved by using a single element transducer (Olympus Immersion Transducers, Waltham, United States), a programmable signal generator (Agilent Technologies 33250A Waveform Generator, Santa Clara, California,

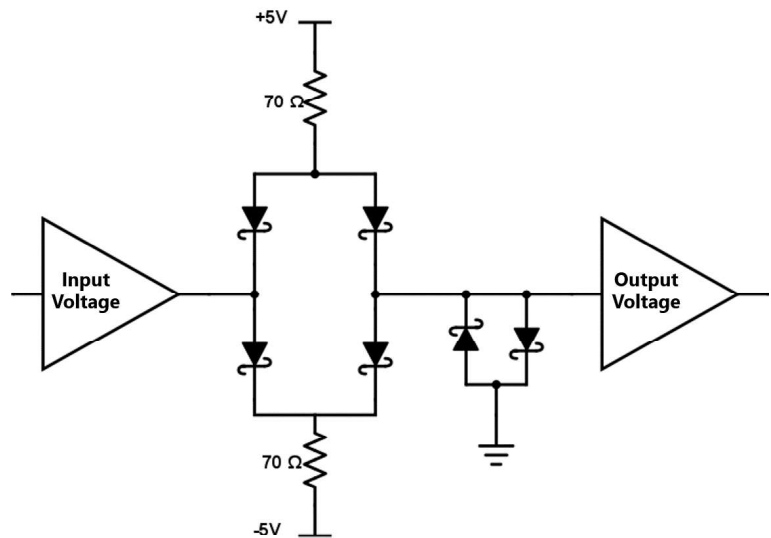

**Figure 3.3:** *Circuit schematic for the Transmit/Receive (T/R) switch, based on the LM96530 module by Texas Instruments* [24]. *The switch was built using Schottky diodes to allow for a switching time short enough to detect both the transmitted waveform and the received echo. High voltage inputs from the amplifier saturated due to the +/- 5V DC supply provided to the T/R switch, whereas low voltage echo signals from the transducer were reproduced at the output of the T/R switch.*

United States), a 55 dB voltage amplifier (A300, E&I Ltd., Rochester, New York, United States), a transmit/receive (T/R) switch, and an oscilloscope (Teledyne LeCroy WaveRunner 104xi, New York, United States). Figure 3.2 outlines how these components were connected to produce the desired signal pathway.

Following the discussion in section 2.4, a pulse inversion protocol was used with this experimental set-up. This was achieved by outputting a 5 cycle sinusoidal pulse from the signal generator and storing the resulting echo in the oscilloscope's memory, at a peak negative pressure of 130 kPa and at varying frequencies. This same procedure was repeated using an identical but inverted pulse ($180^o$ phase shift relative to the non-inverted pulse). The two resulting echoes were then summed using an external computer to generate the residual signal – see section 3.3 for the results of this analysis.

### 3.2.1 Transmit/Receive Switch

One of the challenges of using the same ultrasound transducer to transmit the excitation signal and to also receive the echo was differentiating between the two signals, which travel along the same connection in quick succession. Additionally, it was important to accurately capture the dynamics of the low-voltage received signal to deduce accurate signal energy measurements. A transmit/receive (T/R) switch was therefore designed and fabricated to meet these requirements. Its schematic is depicted in Figure 3.3. This design was based on the Texas Instruments LM96530 Ultrasound T/R switch module [26].The switch saturated high voltage signals (typically between 5 − 10 V after amplification), but reproduced low voltage signals (typically 1 − 50 mV) at its output without any nonlinear distortion. Additionally, the T/R switch prevented damage to the oscilloscope as it saturated the incoming high voltage signal, but allowed the low voltage received echo signal through without saturation. The characterisation of the T/R switch, presented in Figure 3.4, confirmed these properties. As part of the signal chain, the T/R switch functioned correctly to allow the oscilloscope to reproduce and store the low voltage ultrasonic signals scattered by the microbubbles.

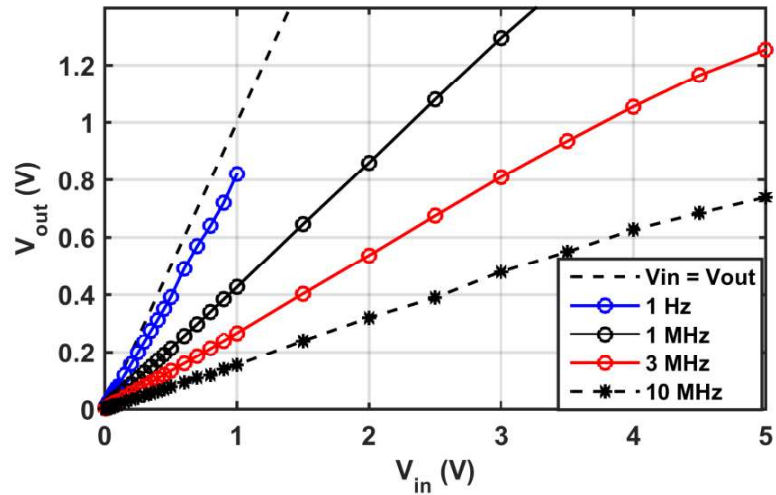

**Figure 3.4:** *The characterisation of the T/R switch with output peak-to-peak voltage plotted as a function of input peak-to-peak voltage (a pure sinusoidal waveform). The switch was tested with input voltage frequencies of 1 Hz (blue line), 1 MHz (black line with circular markers), 3 MHz (red line) and 10 MHz (black line with star markers). The switch behaves linearly, and saturation does not take place below an input 3.5 Vpp (at 3 MHz). Hence, low voltage inputs from the ultrasonic transducer, which are on the order of 1 - 50 mVpp, would be reproduced at the output of the T/R switch without distortion. Furthermore, the sinusoidal waveform itself was not distorted at the output.*

## 3.3   Experiment Results

### 3.3.1 Experiment A – 1.5 MHz driving frequency using a 0.5 MHz centre frequency transducer

The first experiment protocol was designed to generate ten pairs of positive phase and negative phase back scattered echo signals at each hydrostatic pressure level. Three hydrostatic pressure levels were used – no overpressure (absolute atmospheric pressure), 3920 Pa (30 mmHg / 40 cmH$_2$O) overpressure and 6860 Pa (51 mmHg / 70 cmH$_2$O) overpressure. These pressure values were chosen due to the convenience of adjusting the syringe pressurisation system to the respective heights. A SonoVue® microbubble suspension was prepared at a concentration of $10^5$ microbubbles per ml. A 0.5 MHz centre frequency single element ultrasonic transducer (model A301S-SU) was chosen to excite the microbubble solution, spherically focused and driven at its third harmonic at 1.5 MHz so as to lower the Mechanical Index (MI) and reduce the probability of microbubble cavitation [27].

To evaluate the signal content, each recorded signal was cropped in time to contain only the contrast agent echo. This crop time-window was achieved by calculating the time of flight of the ultrasound signal. Figure 3.5 illustrates an example of a raw signal waveform and the cropped version which was low-pass filtered with a 5.5 MHz cut-off to attenuate high frequency noise. For each hydrostatic pressure level, the ensemble average of the set of positive phase signals was taken, as well as the ensemble average of the set of negative phase signals. This resulted in a higher signal-to-noise ratio (SNR) as the presence of random noise was suppressed through averaging. The ensemble average of both the positive and negative phase signals are overlaid in Figure 3.6 – demonstrating that their phase asynchrony should lead to the desired nonlinear residual signal when summed. This analysis relies on the assumption that the system is time invariant: the microbubble properties remain unchanged after being excited once, so that pairing the positive/negative phase responses with each other is valid despite the difference in time between excitations. In reality, some microbubbles would be destroyed after excitation, and although these would be replenished due to the steady flow of solution, there is an underlying assumption that the replenished microbubbles will behave in the same physical manner as the previous generation.

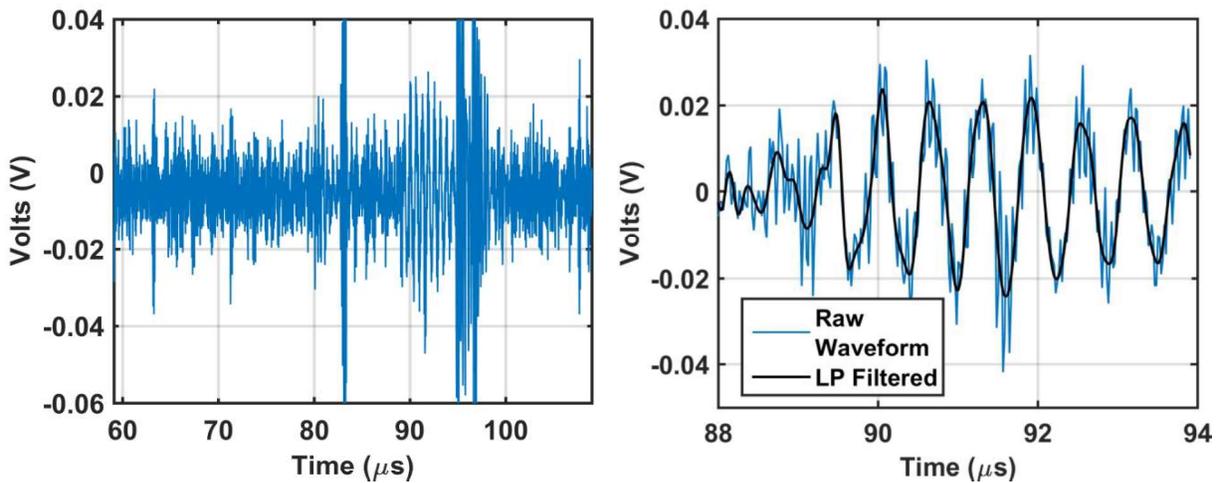

**Figure 3.5:** *Example of a positive phase echo signal as recorded by the oscilloscope through the T/R switch, from an experiment run with exciting frequency of 1.5 MHz and peak negative pressure of 130 kPa.* **Left:** *The raw signal before being cropped. The clock was set to begin counting time with the transmission of the ultrasonic pulse (not shown here). Inspection of the waveform reveals that the contrast agent echo falls between 88 and 94 μs – the time window used for cropping.* **Right:** *The cropped raw and low-pass-filtered signal with the DC offset removed. The blue waveform is the raw echo. The black curve is the raw data after being processed by a low pass Finite Impulse Response (FIR) filter with passband frequency at $f_{PB} = 5.5 \, MHz$, stopband at $f_{SB} = 7 \, MHz$ and a stopband attenuation of $60 \, dB$. The filtered waveform is presented here to clarify the dynamic nature of the contrast agent response.*

In total, the experiment resulted in three pulse inverted residual echo signals, one at each hydrostatic pressure level. Using the analysis described in section 2.4, the energies of each of these signals were calculated. The results are summarised in Table 3.1. The significance of differences in signal energies was judged by analysing the variability of signal energies within each set of hydrostatic pressure levels. This was achieved by creating two groups of five positive and negative phase signals (grouped randomly at each iteration), per pressure level. Using ensemble averaging as above, the residual signal energy of each group was calculated and compared to the other group within the same pressure level. This process was repeated for 1000 iterations to allow the random groupings to achieve many combinations of positive and negative phase signals. This intra-group variation is also presented in Table 3.1. The change in signal energies between pressure level groups was considered significant if its absolute value was greater than the mean of the respective intra-group signal energy variations by more than two standard deviations.

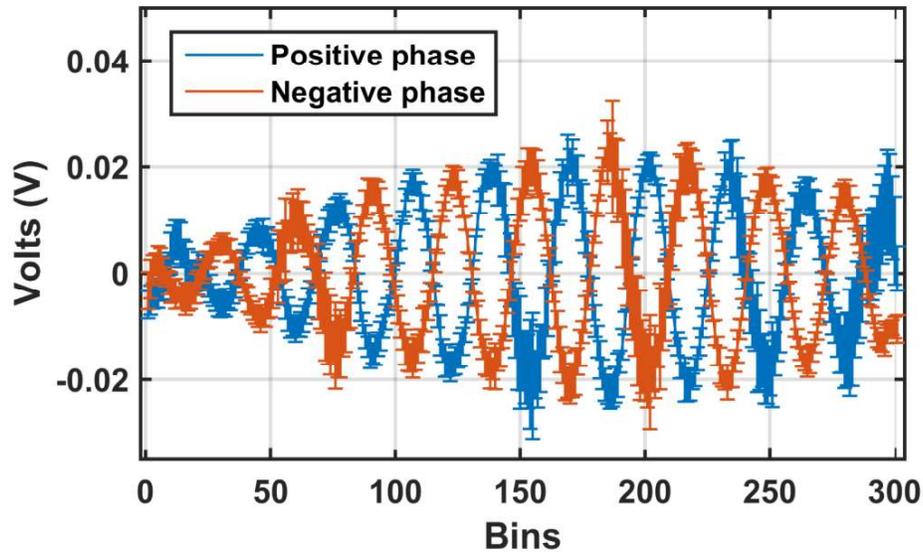

**Figure 3.6:** *Overlay of the ensemble average of echo signals at atmospheric pressure (no overpressure). The blue signal is the average of all echoes from the contrast agent in response to a positive phase $(0^o)$ sinusoidal excitation at 1.5 MHz. The red signal corresponds to the sum of all echoes to a negative phase $(180^o)$ sinusoidal excitation. The error bars plot the standard error of the mean at each point in time for the given phase. The x-axis has been normalised from time into bin values.*

The results of this experiment reveal that there is a significant increase in signal energy when the microbubbles experience a change in ambient pressure from no overpressure to an overpressure of 3920 Pa. Further increasing the overpressure to 6860 Pa did not result in a significant change in signal energy.

| Change in hydrostatic pressure (Percentage change) | Percentage change in PI signal energy | Hydrostatic overpressure level | Mean percentage change in PI signal energies within pressure level | Standard deviation of percentage change in signal energies with pressure level |
|---|---|---|---|---|
| Atmospheric to +3920 Pa (+3.9%) | 410% * | 0 (atmospheric pressure) | 12% | 54% |
| Atmospheric to +6860 Pa (+6.9%) | 440% | 3920 Pa | 24% | 62% |
| +3920 Pa to +6860 Pa (+2.8%) | 6.5% | 6860 Pa | 450% | 560% |

**Table 3.1:** *Summary of results from experiment A. The left two columns present the change in pulse inverted (PI) signal energy with respect to a change in hydrostatic overpressure. An asterisk (*) indicates a change significantly greater than the mean variation within the respective pressure level groups. The right three columns summarise the variations of residual signal energies within each pressure level by presenting the mean and standard deviation of the percentage variation.*

### 3.3.2 Experiment B – 1.75 MHz driving frequency using a 2.25 MHz centre frequency transducer

Following the experiment detailed in section 3.3.1 above, a similar experiment was run to collect more data on signal energy changes caused by hydrostatic pressure changes on the microbubbles. Once again, the signal chain illustrated in Figure 3.2 was used in this experiment. For this experiment a 2.25 MHz single element ultrasonic transducer (model A306S-SU) was used, driven at a frequency of 1.75 MHz. The reason for using a driving frequency different to the centre frequency was to avoid attenuation at the second harmonic content of the response, which plays the biggest role in the pulse inverted residual signal. The bandwidth of the transducer's frequency response is narrow enough that driving at the centre frequency $f_c$ = 2.25 MHz would cause 30 dB of attenuation at the second harmonic and 20 dB of attenuation at the subharmonic. On the other hand, driving at 1.75 MHz would cause 10 dB of attenuation at the second harmonic and 25 dB of attenuation at the subharmonic. The transmitted pulses were sinusoidal pulses of 5 cycles at pre-amplification peak-to-peak voltage of 100 mV. This voltage corresponded to a negative peak pressure of 130 kPa at the focus of the transducer, as before.

| Change in hydrostatic pressure (Percentage change) | Percentage change in PI signal energy | Hydrostatic overpressure level | Mean percentage change in PI signal energies within pressure level | Standard deviation of percentage change in signal energies with pressure level |
|---|---|---|---|---|
| Atmospheric to +3920 Pa (+3.9%) | -16% | 0 (atmospheric pressure) | 9.9% | 45% |
| Atmospheric to +6860 Pa (+6.9%) | 9.5% | 3920 Pa | 8.0% | 43% |
| +3920 Pa to +6860 Pa (+2.8%) | -7.7% | 6860 Pa | 3.1% | 27% |

**Table 3.2:** *Summary of results from experiment B. The left two columns present the change in pulse inverted (PI) signal energy with respect to a change in hydrostatic overpressure. The right three columns summarise the variations of residual signal energies within each pressure level by presenting the mean and standard deviation of the percentage variation. There were no significant changes in signal energy.*

Furthermore, due to the large variations in signal readings observed at 6860 Pa overpressure in the previous experiment, it was decided that 20 pairs of positive and negative phase echo signals should be collected per hydrostatic pressure level to increase the sample size and reduce statistical variance.

The results of the experiment are summarized in Table 3.2, along with the intra-group variations at each pressure level. It can be seen that the changes in signal energy are not large enough to be significant for any of the hydrostatic pressure changes. Increasing the number of samples per pressure group resulted in a reduction of the intra-group variation. Possible reasons for the insignificant changes in signal energies are discussed in section 4.2.2.

### 3.3.3 Experiment C – 3 MHz driving frequency using a 3.5 MHz centre frequency transducer

A final experiment was planned using another single element transducer (model A382S-SU) with a centre frequency of 3.5 MHz being driven at 3 MHz. The purpose of this experiment was to match the driving frequency with that used in the computational model as part of the sensitivity analysis in section 2.5, and hence achieve results more predictable by the numerical model. Unfortunately, running this experiment did not yield any discernible back scattered signals from the contrast agent. On the other hand, the equipment recorded back scattered signals from the empty air-filled channel before the contrast agent solution had been introduced. This discrepancy suggests that the SonoVue® contrast agent solution was not prepared correctly.

## 3.4  Verasonics® Vantage – A Broadband Ultrasound Research System

The experimental model presented in sections 3.2 and 3.3 relied on the use of single element ultrasonic transducers. While these are reliable devices for transmitting ultrasound at particular individual frequencies, such as in High Intensity Focused Ultrasound (HIFU) applications, they are limited by their frequency response range and hence in their ability to receive scattered ultrasound signals at multiple frequencies. As a result, such transducers are not the best tools to use in pulse inversion protocols where it is important to capture harmonic content across a wide range of frequencies. Hence to improve the experimental procedure used above, the Verasonics® Vantage platform (Verasonics Inc., Redmond, WA, USA) was used in a new experimental model, presented in this section. This platform was coupled with the Verasonics® L11-4V transducer, which utilises 128 piezoelectric elements to transmit and receive ultrasound. The transducer has a bandwidth of 4 MHz to 11 MHz. Therefore in the experiment a transient pulse at 4.63 MHz was employed such that the second harmonic at 9.26 MHz lied in the transducer bandwidth, and the residual signal energy contained nonlinear content from the microbubble echoes. The Vantage platform's highly programmable interface to control the pulse transmit characteristics, excitation frequency, peak pressure, beam focusing, capture frequency, and imaging capabilities aided in the design of the experiment model.

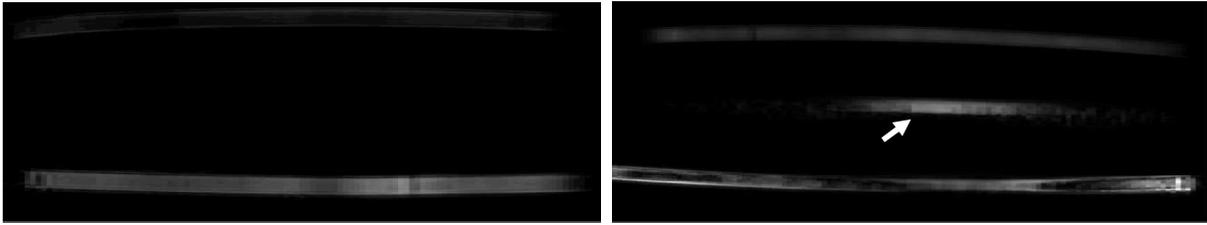

**Figure 3.7**:    *Ultrasound images acquired of the phantom with pulse inversion using the Verasonics® Vantage system. **Left:** Control image of the channel within the phantom filled with water. The two visible impedance boundaries are of the Mylar film which mark the beginning and end of the agar phantom. The channel, which lies between these two films, is not visible. **Right:** Image of the phantom's channel when filled with a SonoVue® microbubble suspension at 2 % concentration by volume. The microbubbles, shown by the arrow, are clearly visible at the centre of the phantom, as they strongly scatter the ultrasound signal and illuminate the channel.*

The Verasonics® Vantage platform included built-in modules which did not require the signal generator, amplifier, T/R switch and oscilloscope as with the previous signal chain. However, the need to suspend and pressurise the contrast agent solution still existed, and so the same phantom design and pressurisation system detailed in section 3.2 were used. Furthermore, SonoVue® microbubbles were used once again as the contrast agent suspension.

The pulse inversion protocol was programmed such that the transducer would transmit a positive phase sinusoidal pulse of 5 cycles at a frequency of 4.63 MHz and peak negative pressure of 100 kPa. After transmitting, the transducer would prepare to receive the resulting echo signal. After receiving, the transducer was programmed to automatically transmit an inverted pulse of the same characteristics, and then receive the resulting echo. Hence, the Vantage system could perform instantaneous pulse inversion through the summation of the resulting echoes. This contrasts with the previous experimental model, where it was necessary to manually invert the phase before transmission.

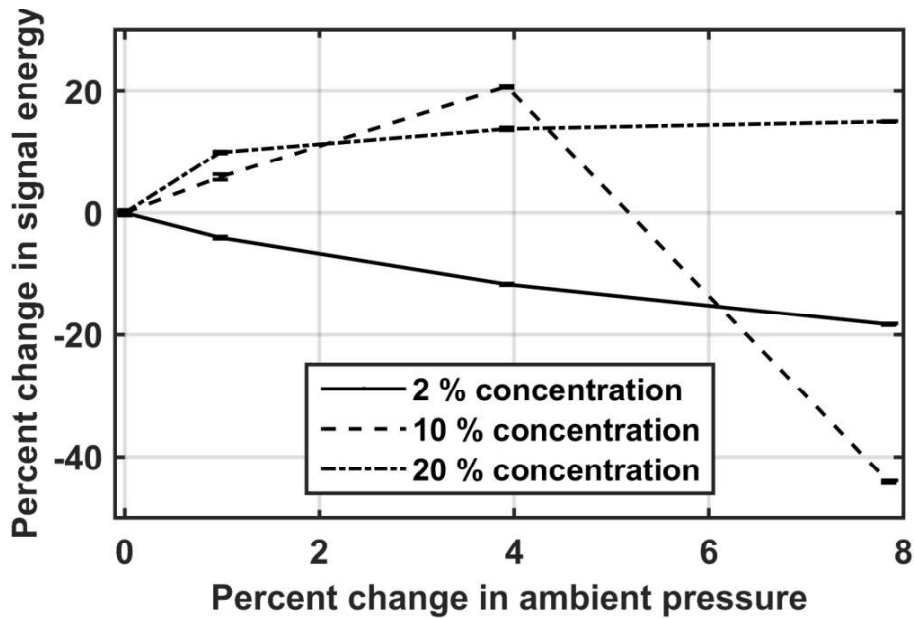

**Figure 3.8:** *Results of the Verasonics® experiment, showing the percentage change in average signal energy caused by a change in ambient pressure. The pressure was increased from atmospheric pressure (zero overpressure) to 980 Pa overpressure (0.98 % increase), to 3920 Pa overpressure (3.9 % increase), to 7850 Pa overpressure (7.9 % increase). Each change was significant, as verified by paired sample t-tests. The experiment was carried out at three different concentrations of SonoVue® contrast agent. Error bars plot the standard error of the mean of the measurements.*

The experiment was performed with SonoVue® suspensions at three different concentrations: 2%, 10% and 20% volume of SonoVue® diluted in deionised water. The contrast agent was infused through the channel at 0.2 µl/min. For each concentration, four hydrostatic pressure levels were tested: atmospheric pressure (no overpressure), 980 Pa (7.4 mmHg / 10 cmH$_2$O) overpressure, 3920 Pa (30 mmHg / 40 cmH$_2$O) overpressure, and 7850 Pa (59 mmHg / 80 cmH$_2$O) overpressure. Furthermore for each pressure level, ten pulse inverted residual signals were recorded in order to generate enough samples to make statistical inferences.

Figure 3.7 compares the control scenario – where the phantom's channel was infused with pure deionised water – with the case of a steady infusion of microbubbles through the channel. The presence of the microbubbles as a bright region in the centre of the channel can be observed.

Figure 3.8 shows the measured change in signal energy as a function of overpressure. For the each concentration of SonoVue®, the signal energy was determined by summing the array values of the pulse inversion intensities from regions in the channel. For a 2% concentration of SonoVue® there was a monotonous decrease in signal energy with overpressure. For 10% and 20% SonoVue® concentrations the signal energy increased with overpressure. However for a 10% concentration, there was a large decrease in signal energy at a 7.9% increase in overpressure. Paired sample t-tests were performed to test the significance of each of the changes in signal energy; each change was significantly different ($p < 0.05$).

Although significant changes in signal energy were detected with increasing ambient pressure, the resulting trends at the different contrast agent concentrations are not consistent with each other. Discussion of this limitation of the results is presented in section 4.1.

# 4. General Discussion and Conclusion

## 4.1 Implications of Experiment Results

The results of the four experiments detailed in section 3 carry a range of implications. Firstly, the significant change in residual signal energy due to a change in ambient pressure observed in experiment A has bearings on the validity of the theoretical models presented in section 2. In particular, the experiment confirmed that there is indeed a change in the dynamic behaviour of microbubbles with a change in ambient pressure, as predicted by both the RPNNP and Marmottant models. More importantly, observing this change has shown the feasibility of using microbubbles, and standard ultrasound and signal processing equipment, to observe a change in a fluid's ambient pressure. This result is important as it provides a basis to carry out further investigations into the exact nature of this trend, which could see the verification of the proposed model and trend in section 2.5. However, the variability caused by noise in some of the recorded signals (see section 4.2.1) suggests that the process needs to be further refined such that it is possible to reliably measure changes in signal content and hence in a clinical setting be able to monitor the changes in a patient's blood pressure.

On the other hand, the contrasting results of experiment B imply that it is not feasible to detect a change in hydrostatic pressure, as there was no significant change in pulse inverted signal energy. There are a few explanations as to why experiments A and B would produce these opposing results. One explanation is that the difference in initial conditions of the system – explicitly the different driving frequencies and transducer models – were significant enough to change the response of the underlying model and hence the observed response. However given the limited scope of the experimental model (as defined in section 3.1) it is difficult to quantify the complete difference in initial conditions as there is no control or measure on all the model parameters. Hence, it is difficult to discern whether the difference in results between experiments A and B is caused by the lack of control over model parameters or because of the microbubbles being unaffected by a change in ambient pressure.

The experiment performed with the Verasonics® Vantage system (section 3.4) produced results that also strongly suggest that detecting a change in ambient pressure is practically achievable using ultrasound and microbubbles. Furthermore, a clearer signal was achieved compared to the former experimental model. This can be explained by the instantaneous pulse inversion method employed by the system software, in addition to the broadband frequency response of the associated transducer. Thus, the Verasonics® model will be a good candidate for future experimental research in this field, and perhaps for clinical application, as it is capable of producing more reliable results.

In contrast, the results produced by the Verasonics® experiment were limited in that they did not produce consistent trends when repeated with different concentrations of SonoVue® suspensions. While the 2% concentration suspension produced a signal energy trend that monotonically decreased with increasing pressure, the 20% suspension concentration produced a monotonically increasing trend, and the 10% concentration suspension did not produce a monotonic trend. Based on the linear population assumption (from section 2.6), it would be expected that the resulting trends in signal energy changes would be similar for any concentration of contrast agent. However, at higher contrast agent concentrations the possibility of nonlinear multiple scattering is increased, where the assumption of individual microbubbles in an infinite medium breaks down [24]. Another possible explanation is that various model parameters changed over time and hence changed the dynamics of the system. For instance it is possible that in the time taken to run the experiments with different concentrations, properties such as the surface tension of the microbubbles' coatings had changed by an amount to significantly affect the response of the microbubbles. This could also be possible with parameters such as the polytropic exponent of the encapsulated gas (sulphur hexafluoride in SonoVue® [28]), which can change with respect to time due to gas diffusion out of the microbubble. It is recommended that in future experiments signal energy measurements should be repeated at no overpressure to observe the effects of time on the model parameters. Such time sensitive effects should be taken into account to improve the existing model.

## 4.2 Limitations of the Experimental Model

The experiments presented in sections 3.3.1, 3.3.2, 3.3.3 could not produce results that were wholly consistent with each other or with the mathematical model. For instance, the significant change in pulse inverted signal energy observed in experiment A was present only in one of the instances of the hydrostatic pressure change. Moreover, such a significant change was not observed again in experiment B. Finally, experiment C was not consistent with respect to the other experiments as it could not produce discernible echo signals from the contrast agent solution. This section will discuss the possible causes for these inconsistencies, and how they can be addressed in future studies.

### 4.2.1 Excessive Noise Artefacts

The echo signals recorded in experiment A at an overpressure of 6860 Pa varied significantly with respect to pulse inverted signal energy (see Table 3.1). As a result, it was not possible to conclude whether the observed percentage increase in signal energy of 440% between no overpressure and 6860 Pa overpressure was significant. This was in contrast to the signal readings at no overpressure and 3290 Pa of overpressure, which did not show such large variations. It is possible to understand the cause of these variations by studying the individual received echoes.

Figure 4.1 illustrates an example of a typical noisy echo signal received at 6860 Pa overpressure. It can be seen that there are noise artefacts that cannot be easily removed through low pass filtering as before with random high frequency noise. These noise artefacts occur in bursts and are high amplitude signals. The frequency of these bursts is higher than the transducer's driving frequency but lower than the low-pass filter's cut-off, and so low-pass filtering does not recover the underlying low frequency trend caused by microbubble oscillations. A filter with a lower passband frequency to attenuate the noise artefacts is not used as this might also attenuate any content at the second harmonic, which is crucial in capturing the nonlinear dynamics of the microbubbles. Furthermore, the effects of these noise artefacts are not readily minimised by taking the ensemble average of the signals. Therefore, the resulting residual signals will also contain a significant proportion of this noise, and any signal energy calculations will not be able to distinguish between this noise and the desired contrast agent signal. Due to the frequent random occurrence of these noise artefacts within the echo signals at 6860 Pa overpressure, the overall energy response varied significantly.

The fact that the noise artefacts did not appear at the same time in each signal, suggests that they are not caused by any physical structure in the path of the ultrasound beam. It is more likely that they are caused by electrical interference from the surrounding laboratory equipment, either through a direct connection or by electromagnetic interference. In addition to recording more samples of contrast agent echoes (as in experiment B) to reduce the proportion of noisy signals, it would be advisable to ensure each module within the signal chain is decoupled and isolated from surrounding equipment.

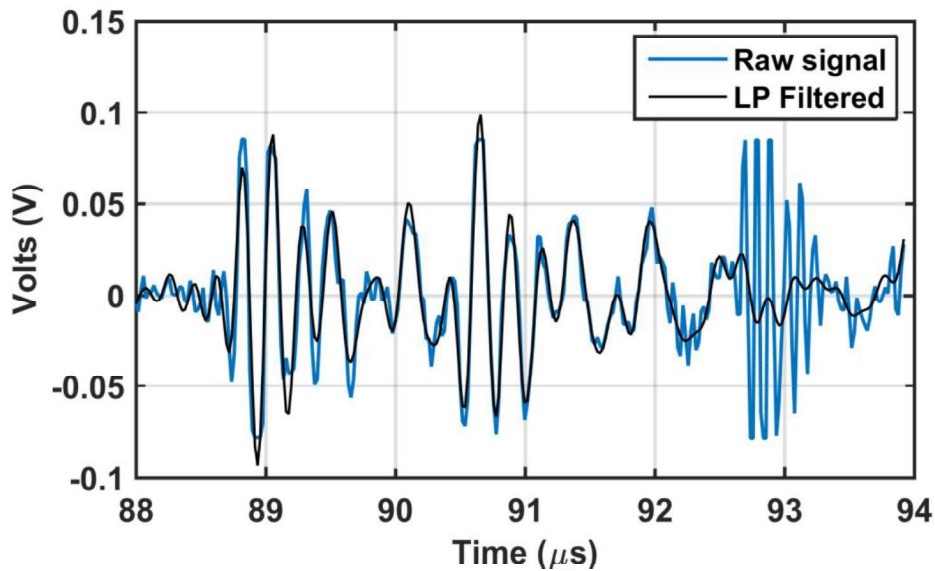

**Figure 4.1:** *An example of a raw signal (in blue) containing significant noise artefacts at 88.8 μs, 90.4 μs and 92.6 μs. The low pass (LP) filtered signal is in black. The first two bursts are not attenuated by the LP filter. The final burst contains frequencies high enough to be attenuated by the LP filter, but its high amplitude masks the underlying oscillations caused by the microbubble echo. The microbubbles were excited with a 1.5 MHz sinusoidal pulse. The filter is a FIR filter with passband frequency at $f_{PB} = 5.5\ MHz$, stopband at $f_{SB} = 7\ MHz$ and a stopband attenuation of $60\ dB$.*

### 4.2.2 Narrowband Transducer Frequency Response

As mentioned in section 3.4, one of the limitations of experiments A, B and C was the use of single element transducers with relatively narrowband frequency responses. It is possible that despite selecting driving frequencies different to the centre frequencies of the transducers, the attenuation at the subharmonic and second-harmonic frequencies was too great. Although microbubbles oscillate nonlinearly, their response at frequencies other than the fundamental is smaller than at the fundamental frequency (see section 2.6). Hence, the expected echo signal will contain nonlinear content at relatively low voltages (on the order of a 1 mVpp). This nonlinear content would be then attenuated due to the transducer's frequency response which would further drive down the voltage level, and hence decrease the SNR at the desired frequencies. This is a possible explanation as to why the results of experiment B did not show any significant changes in the residual signal energy when hydrostatic pressure was varied. It is probable that the level of noise exceeded the desired signal content at the nonlinear frequencies.

## 4.3  Future Research and Development

With the aim of maximising clinical impact for patients, a reliable method of non-invasively detecting a change in ambient pressure must be developed. The work in this report has provided proof-of-principle, but has been limited experimentally in that it was not within the scope to reliably verify the trends predicted by the mathematical model. The trends, if reproduced *in vitro*, *in vivo* and clinically, would hold great importance in medical diagnostics, as discussed in section 1.1.

The first step in this process would be to carry out further experiments with the Verasonics® system and SonoVue® contrast agent, to see if the numerical model's predicted signal energy trends could be verified. To achieve this the numerical model must be initialised and run with the estimated parameter values of the experiment, such as the appropriate driving frequency, and the estimated mean microbubble radius and initial surface tension. Confirming the model with a clinically available contrast agent would increase the potential clinical impact of the procedure.

However, if SonoVue® does not produce consistent trends in signal energy, another approach must be adopted to verify the numerical model. In order to achieve this, methods to measure and control the model parameters must be developed. Such methods would be used to design *in vitro* experiments in which the exact state of the microbubble population would be known and monitored. This would allow the calculation of the sensitivity of the microbubbles' response to changes in parameters. Parameters such as the initial microbubble radius $R_0$ can be measured and controlled well using existing methods – microfiltration and centrifuging, for example [5]. On the other hand, further work will have to be done to develop effective ways of quantifying parameters such as the shell elasticity $\chi$ and shell viscosity $\kappa_S$. Once it is possible to measure and control these parameters, the numerical model can also be formally optimised under constraints to produce an optimal set of parameter values which can be used to maximise the sensitivity of the response to changes in ambient pressure, while minimising the sensitivity to the other parameters. This optimal parameter set can then be implemented in the experimental work. In clinical applications, a more practical method to achieve this would be to calibrate the response of the microbubbles by measuring their behaviour under a known relative change in hydrostatic pressure – for example between different blood circulation sites in the body.

## 4.4  Conclusions

This project aimed to quantify the feasibility of using ultrasound and microbubbles to measure changes in the ambient pressure of fluid. The dynamic behaviour of microbubbles under ultrasonic excitation was numerically modelled using existing time-frequency analytical techniques in ultrasound. The numerical model also examined the effect of changing ambient pressure on the response of microbubbles. The simulations suggested that the pulse inverted signal energy from a microbubble should decrease by 11% for an ambient pressure increase of 1%. The numerical model was used to quantify the extent to which variations in microbubble and ultrasound properties would cause the theoretical dynamic response to change. The analysis yielded that a microbubble's response was most sensitive to changes in ambient pressure and initial bubble radius, with respect to signal energy measurements. The model was extended to calculate the response of a population of heterogeneous microbubbles. Overall, the computational model indicated the theoretical feasibility of using microbubbles to measure a change in ambient pressure as low as a 1% increase of atmospheric pressure.

To study the feasibility of practically measuring a change in ambient pressure, experiments were designed and carried out using two different ultrasound systems. Based on the metric of pulse inverted signal energy, both of these experiments independently verified that it is possible to detect a change in the ambient pressure acting on the microbubbles, with the Verasonics® model showing that a 0.98% increase in ambient pressure is detectable. Finally the limitations of the work were presented, along with possible methods of improving the numerical and experimental techniques to aid future research. Therefore, the aims of the project were satisfied both in a theoretical and practical nature.

# Appendix A – MATLAB Programs

**MATLAB function representing the Marmottant model (from section 2.2)**

```matlab
function a = lrp(t,r)

% Global model parameters
global po sigma1 sigmaRo Ro1 k amp mul f rhol Rbuck Rrup chi ks tspan inv_flag

w = inv_flag*gauss_pulse(t,tspan(end)/2,tspan(end)/6);    % Gaussian enveloped
forcing function with mean tmax/2 and SD = tmax/6

% Buckling and rupturing surface tension criteria
if r(1) >= Rbuck && r(1) <= Rrup
    sig = chi*((r(1)./Rbuck).^2 - 1);
elseif r(1) > Rrup
    sig = sigma1;
else
    sig = 0;
end

% Differential equation describing microbubble dynamics - Marmottant model
a = [r(2);...
    (((po+(2.0*sigmaRo/Ro1))*Ro1^(3*k)*(r(1)^(-3*k))) ...
    -po ...
    -amp*w ...
    -2.0*sig*(r(1)^(-1))...
    -4.0*r(2)*mul*(r(1)^(-1))...
    -4.0*ks*r(2)*(r(1)^(-2)) ...
 )/rhol ...
    -1.5*(r(2)^2) ...
    )*r(1)^(-1) ...
];

end
```

**Script to calculate trend in signal energy due to changing hydrostatic pressure, taking into account a microbubble population with varying initial radii (from section 2.6)**

```matlab
% Checks how a population of bubbles responds to excitation, in particular
% bubbles' radius (Ro) distributed normally
%% Initialisation of parameters
global po po2 sigma1 sigmaRo Ro1 k amp mul f rhol Rbuck Rrup chi ks tspan
inv_flag

dpo = 1.0e5;             % ambient pressure (Pa)
dpo2 = linspace(1e5, 1.2e5, 20);  % Ambient pressure after compression (Pa)
dsigma1 = 0.07;         % equilibrium gas/shell surface tension (N/m)
dsigmaRo = 0.03;
dmul = 1.0e-3;          % surrounding fluid viscosity (Pa.s)
dchi = 1;               % Elasticity of coating (N/m)
drhol = 1000.0;         % surrounding fluid density (kg/m^3)
dco= 1498;              % surrounding fluid phase velocity (m/s)
dks = 2.3e-8;           % effective surface viscosity (Ns/m)
k = 1.4;                % ratio of gas specific heats

dRo1_mean = 3e-6;       % Mean of distribution of bubbles' radius
dRo1_sd = 0.3e-6;       % Standard deviation of distribution of bubbles' radius
```

```matlab
N = 100;                % Number of bubbles in population

% Initial radius at r=Ro - distributed normally. Each row
% corresponds to an increasing SD of sigmaRo. Columns represent the 100
% distributed values
dRo1 = normrnd(dRo1_mean*ones(length(dRo1_sd),N),repmat(dRo1_sd',1,N), ...
[length(dRo1_sd) N]);

dRbuck = dRo1.*((dsigmaRo./dchi)+1).^-0.5;  % bubble bucking radius (m)
dRrup = dRbuck.*(1+(dsigma1/dchi))^0.5;     % bubble rupturing radius

dRbuck_mean = dRo1_mean*((dsigmaRo/dchi)+1).^-0.5; % bubble bucking radius (m)
dRrup_mean = dRbuck_mean*(1+(dsigma1/dchi))^0.5;   % bubble rupturing radius

df = 3e+06;          % insonation frequency (Hz)
n = 5;               % no. cycles
dtmax = n/df;        % run time
damp = 50e3;         % insonation pressure amplitude (Pa)
tmax = dtmax*2*pi*df;
tspan = linspace(0,tmax,5000);
dt = tspan(2)-tspan(1);    % regular time step
Fs = 1/dt;                 % sampling frequency
dFs = 2*pi*df*Fs;          % dimensional sampling frequency

%% Solving Marmottant model for each microbubble in each population (varying
with spread)
for j2 = 1:length(dRo1_sd)
    for j3 = 1:length(dpo2)
        for j = 1:size(dRo1,2)
            radiated2 = 0;
            for flag = 1:2
                % Non-dimensionalising parameters
                po = dpo/damp;
                po2 = dpo2(j3)/damp;
                rho1 = drho1*(dRo1_mean*2*pi*df)^2/damp;
                mul = dmul*df*2*pi/damp;
                Ro1 = dRo1(j2,j)/dRo1_mean;
                Rbuck = dRbuck(j2,j)/dRo1_mean;
                Rrup = dRrup(j2,j)/dRo1_mean;
                sigma1=dsigma1/(damp*dRo1_mean);
                sigmaRo = dsigmaRo/(damp*dRo1_mean);
                chi = dchi/(damp*dRo1_mean);
                ks = (dks*2*pi*df)/(damp*dRo1_mean);
                f = df/(2*pi*df);
                amp = damp/damp;

                % Calculate static changes in microbubbles due to change in
                % hydrostatic pressure
                if po ~= po2
                    R2 = fsolve('overpressure',Ro1);
                    Ro2(j3,j,j2) = R2;
                    dRo2(j3,j,j2) = Ro2(j3,j,j2)*dRo1_mean;
                    dsigmaRo2(j3,j,j2) = ...
lap_p(dRo2(j3,j,j2),dchi,dRbuck(j2,j),dRrup(j2,j),dsigma1);
                    sigmaRo2 = dsigmaRo2(j3,j,j2)/(damp*dRo1_mean);
                    Ro1 = Ro2(j3,j,j2);
                    sigmaRo = sigmaRo2;
                    po = po2;
                end

                % Apply inverted pulse for every second iteration
                if mod(flag,2) == 0
```

```matlab
                    inv_flag = -1;
                else
                    inv_flag = 1;
                end

                % Scattered pressure equation
                pr =@(r,v,a,R) rhol*((1./R).*((r.^2).*a + 2*r.*v.^2) - ...
((r.^4).*v.^2)./(2*R.^4));
                % Numerically solve DE
                t = 0.0;
                r = 0.0;
                try
                    % if error occurs, ignores it and skips this iteration
                    [t,r] = ode45('lrp',tspan,[Ro1;0]);
                    tm = (t*1e6)/(2*pi*df);
                    rad = r(:,1);       % radius non-d
                    vel = r(:,2);       % velocity non-d
                    % acceleration non-d:
                    acc = gradient(r(:,2),tspan(2) - tspan(1));
                    scatp = pr(rad,vel,acc,10*Ro1); % Scattered pressure non-d
                    % Residual sum of pressures for each bubble:
                    radiated2 = radiated2 + scatp;
                    % Scattered pressure for total population:
                    if flag == 2
                        total_dist_pressure(j3,:,j2) = ...
                        total_dist_pressure(j3,:,j2) + radiated2';
                    end
                catch
                    continue
                end
            end
        end
    end
end

%% Calculate frequency spectrum
for j2 = 1:length(dRo1_sd)
    for j3 = 1:length(dpo2)
        L1 = length(rad);
        N1 = 2^(nextpow2(L1)+5);     % For zero padding
        X = fft(detrend(rad),N1);    % Fourier transform of radius-time curves
        P1 = abs(X/L1);
        P1 = P1(1:(N1/2)+1);
        P1(2:end-1) = 2*P1(2:end-1);
        P1p = P1.^2;                 % Amplitude squared
        freq1 = ((1/(tm(2)-tm(1)))*(0:N1/2))/N1;   % Frequency scale

        %%% FFT to analyse pulse inverted sum - emphasizes nonlinearities
        L3 = length(total_dist_pressure(j3,:,j2));
        N3 = 2^(nextpow2(L3)+5);
        Y3 = fft(detrend(total_dist_pressure(j3,:,j2)),N3);
        P3 = abs(Y3/L3);
        P3 = P3(1:(N3/2)+1);

        % Amplitude Spectrum of scattered pressure
        P3(2:end-1) = 2*P3(2:end-1);
        P3p = abs(Y3).^2;
        P3p = P3p(1:(N3/2)+1);
        P3p(2:end-1) = 2*P3p(2:end-1);  % Energy Spectral density

        % Total Signal Energy
        energy_PI(j3,j2) = sum(P3p)*N3^-1*dt;
        % Total Signal Energy - check as per Parseval's theorem
```

```matlab
        energy2_PI(j3,j2) = sum(abs(total_dist_pressure(j3,:,j2)).^2)*dt;

        % Find index of relevant frequencies
        subharmonic_freq = find(abs((df*0.5*1e-6)-freq1)<0.008,1);
        harmonic_freq = find(abs((df*2*1e-6)-freq1)<0.005,1);
        fund_freq  = find(abs((df*1e-6)-freq1)<0.01,1);

    end
end

%% Calculate percentage changes in signal energy and hydrostatic pressure
% For each population
for j2 = 1:length(dRo1_sd)
    percentage_energy(:,j2) = ((energy_PI(:,j2) - ...
     energy_PI(1,j2))./energy_PI(1,j2)) * 100;
end
percentage_pressure = ((dpo2(:)-dpo2(1))./dpo2(1)) * 100;
```